# Modeling of flow generated sound in a constricted duct at low Mach number flow

2017

Peshala Thibbotuwawa Gamage
*University of Central Florida*

MODELING OF FLOW GENERATED SOUND IN A CONSTRICTED DUCT
AT LOW MACH NUMBER

by

PESHALA THIBBOTUWAWA GAMAGE
B.S University of Moratuwa, 2013

A thesis submitted in partial fulfillment of the requirements
for the degree of Master of Science
in the Department of Mechanical and Aerospace Engineering
in the College of Engineering and Computer Science
at the University of Central Florida
Orlando, Florida

Fall Term
2017

Major Professor: Hansen A. Mansy

# ABSTRACT

Modelling flow and acoustics in a constricted duct at low Mach numbers is important for investigating many physiological phenomena such as phonation, generation of arterial murmurs, and pulmonary conditions involving airway obstruction. The objective of this study is to validate computational fluid dynamics (CFD) and computational aero-acoustics (CAA) simulations in a constricted tube at low Mach numbers. Different turbulence models were employed to simulate the flow field. Models included Reynolds Average Navier-Stokes (RANS), Detached eddy simulation (DES) and Large eddy simulation (LES). The models were validated by comparing study results with laser doppler anemometry (LDA) velocity measurements. The comparison showed that experimental data agreed best with the LES model results. Although RANS Reynolds stress transport (RST) model showed good agreement with mean velocity measurements, it was unable to capture velocity fluctuations. RANS shear stress transport (SST) k-ω model and DES models were unable to predict the location of high fluctuating flow region accurately.

CAA simulation was performed in parallel with LES using Acoustic Perturbation Equation (APE) based hybrid CAA method. CAA simulation results agreed well with measured wall sound pressure spectra. The APE acoustic sources were found in jet core breakdown region downstream of the constriction, which was also characterized by high flow fluctuations. Proper Orthogonal Decomposition (POD) is used to study the coherent flow structures at the different frequencies corresponding to the peaks of the measured sound pressure spectra. The study results will help enhance our understanding of sound generation mechanisms in constricted tubes including biomedical applications.

# ACKNOWLEDGMENTS

I am mostly grateful to Dr. Hansen Mansy, my thesis advisor, who has inspired me to work on this topic and introduced me to field of Bio-acoustics engineering. I would also like to thank Dr. Alain Kassab and Dr. Samik Bhattacharya, whom have supported me by their valuable advices to become more confident on my research topic.

I would also like to take the opportunity to thank my family and friends for their support throughout my education.

# TABLE OF CONTENTS

# LIST OF FIGURES

# LIST OF TABLES

# CHAPTER 1: INTRODUCTION

Related to many industrial and biomedical applications, the study of flow and acoustics in a constricted duct at low Mach numbers represent a topic of interest. A proper understanding of the mechanisms behind sound generation in such applications will help reduce noise in industrial applications or diagnose pathologies which lead to serious medical conditions. Precise numerical modeling has proven to accurately predict flow generated acoustics and identify the sources that they originate from. This study denotes the validation of a numerical modeling approach to simulate the phenomenon of flow generated sound in the previously listed applications.

Flow in a constricted duct has been previously studied both numerically and experimentally where the focus is mostly given to the study of change in velocity, pressure and shear forces to find effect on flow and vessel wall [1-3]. Furthermore, some studies have carried out detailed studies of the jet-like flow structure generated by the sudden constriction of the flow, concentrating on the recirculation and reattachment regions [4, 5]and the turbulence mixing behind the constriction[6].

Certain studies have focused on flow generated acoustics in a constricted pipe. Most of these studied the wall pressure fluctuations downstream the constriction and discussed the relation between the spectral peaks of the wall pressure fluctuations and turbulent flow dynamics caused by the velocity jet generated at the constriction. The study by Sanaa [7] analyzed velocity fluctuations in jet centerline and the shear layer region together with pressure fluctuations on the wall and concluded that the wall pressure fluctuations are imposed by a passage turbulent eddies generated by the turbulent jet. The studies by Borisyuk [8, 9] have experimentally measured the wall pressure fluctuations and found that the power spectrum of the wall pressure fluctuations behind the narrowing had isolated frequency peaks and these peaks varied with the increment of constricted level. Also, these studies discussed the relations between the characteristic frequencies of the large-scale eddies developed behind the constriction and the frequency peaks in the wall pressure spectrum.

The same problem has been investigated in perspective of flow generated acoustics in a stenosed artery [10-12]. The early study by Bruns [11] suggested that arterial bruits are unlikely generated due to the post-stenotic turbulence considering the low strength quadrupole sources in low Mach number flow. Later, Dewey [13] showed that there is a striking similarity between the spectrums produced by turbulent flow in a pipe and the arterial sounds produced by a stenosis by comparing the spectrum of the recorded arterial bruits and the wall pressure spectrum of fully developed turbulent pipe flow. This study suggested that the flow through a stenosis produces a jet flow which becomes unstable even at low Re numbers and converts portion of its kinetic energy in to turbulent fluctuations which produces pressure fluctuations at the vessel wall which are detected as arterial sounds.

Zhao[14] simulated sound generated from a confined pulsating jet for a low Mach number (0.2) flow, with 2D compressible flow simulation using Lighthill's analogy and direct noise calculation (DNC). Here, the assumption was made that the sound waves travels axially and reflections are ignored. The study found good agreement with the far field acoustic pressure in downstream to the stenosis solved based on Lighthill's analogy and DNC. Based on Lighthill's analogy the study presented three main sound generating mechanisms including, monopole source caused by fluctuating volume velocity, dipole source caused by unsteady forces exerted on duct walls and quadrupole source caused by the presence of vortex pairing.

However, these conclusions contradict with Bruns [11] who argued that the strength of turbulent sources are exceedingly weak for low Mach numbers. Hardin and pope [12] simulated sound generation by a stenosed pipe, employing a 2D geometry of a circular duct with sharp constrictions. They used an acoustic/viscous splitting method suggested by Hardin and Pope [12] to simulate the acoustics. Employing the assumptions that the acoustic waves propagates axially behind the stenosis and ignoring the sources on the solid boundaries, the results showed a good agreement with the experimental spectral data at the outlet of a stenosed pipe.

Gloerfelt[15] used direct numerical simulation (DNS) to simulate the acoustics behind a rectangular duct with a sharp constriction at low Mach number flow. The study concluded the breakdown of the coherent jet-column structure behind the constriction is responsible for the most part of the acoustic energy and small scale of turbulence is contributing in the broadening the spectral response. Seo [16] numerically simulated the generation and propagation of acoustics from a stenosed artery using incompressible flow simulation to simulate flow dynamics and Linearized perturbed compressible equations (LPCE)[17] to simulate acoustics. This study concluded that the primary source of arterial bruits is the vortex inducted perturbations in the near post-stenotic region and acoustic fluctuation induced by the blood flow has a stronger intensity and higher frequency content for the higher level of constriction.

## 1.1 Research Objective

The main focus of this work is to use computational tools to accurately simulate the flow field and flow induced acoustics in biomedical applications such as lung airways and arteries. Flow fields in these types of applications are confined (internal flow) with relatively low Mach numbers. Based on the dimensions of such geometries, acoustic field can be considered as a near-field problem. To investigate flow field and flow induced acoustics in such conditions, a simple geometry of a constricted pipe is selected. First, the flow field is modeled using CFD, employing different turbulence models. The flow results are validated against LDA velocity measurements to select the best turbulence model. Then, the flow induced acoustic field is modeled using a CAA model in parallel with CFD simulation. The results from the CAA simulation are validated against the sound pressure spectra measured on the pipe wall. The acoustic sources and propagation are numerically investigated. Finally, the validated CFD and CAA methods are applied to simulate the flow and flow induced acoustic field in a realistic lung airway geometry.

# CHAPTER 2: METHODS

## 2.1 Modelling of Flow

For the current application flow can be considered as incompressible due to the low Mach number flow and it's also assumed that flow is isothermal due to negligible changes in temperature of the flow. Based on these assumptions, the governing equations for the flow are presented as following.

$$\frac{\partial u_i}{\partial x_i} = 0 \tag{1}$$

$$\rho \frac{\partial u_i}{\partial t} + \rho u_j \frac{\partial u_i}{\partial x_j} = -\frac{\partial p}{\partial x_i} + \mu \frac{\partial^2 u_i}{\partial x_j \partial x_j} \tag{2}$$

Here, Equation 1 and 2 represent the mass conservation and momentum conservation respectively. Subscripts i and j denote the Cartesian tensor notations where the repeated subscripts denote summations over the 3 coordinates. In equation 1 & 2 , *u, p* and *t* are the three-dimensional velocity vector and static pressure and time, respectively. The fluid properties were such that: the density, $\rho$ = 1.2 kg.m-3 and dynamic viscosity, $\mu$= 1.85×10-5 kg.m/s.

As the objective of the current study is to model flow generated sound, the choice of turbulence model is very important. Hence, different turbulent models were used to validate the CFD results comparing with the velocity measurements using Laser Doppler Anemometry (LDA) and then the appropriate model was used for the aeroacoustics simulation.

### 2.1.1 Reynolds Averaged Navier Stokes (RANS) Models

RANS equations are derived by applying the Reynolds decomposition to the flow variables in Navier-Stokes equations. For a flow quantity , Reynolds decomposition is applied as following.

$$\varphi = \langle\varphi\rangle + \varphi' \qquad (3)$$

Here, $\langle\varphi\rangle$ represents the mean value or the ensembled average of the flow quantity while $\varphi'$ and $\varphi$ represents the fluctuating and instantaneous terms, respectively. After applying the Reynolds decomposition to flow variables, RANS equations are obtained.

$$\frac{\partial \langle u_i \rangle}{\partial x_i} = 0 \qquad (4)$$

$$\frac{\partial \langle u_i \rangle}{\partial t} + \frac{\partial \langle u_i \rangle \langle u_j \rangle}{\partial x_j} = -\frac{1}{\rho}\frac{\partial \langle p \rangle}{\partial x_i} + \upsilon \frac{\partial^2 \langle u_i \rangle}{\partial x_j \partial x_j} - \frac{\partial \langle u_i' u_j' \rangle}{\partial x_j} \qquad (5)$$

where, $\upsilon$ is the kinematic viscosity. In Equation 5, the first term in left hand side represents the mean momentum change in a fluid element due to the unsteadiness in the mean flow while the second term describes the mean momentum change due to convection by the mean flow. These momentum changes are balanced by the source terms in the right-hand side which consist of mean pressure, viscous stresses and the third source term which contains the fluctuating velocity components. The term $\langle u_i' u_j' \rangle$ is known as "Reynolds stress" and requires additional modelling to solve RANS equations.

Equation 5 can be further simplified using the turbulent viscosity relation introduced by Boussinesq [18] also known as Boussinesq approximation. Boussinesq described that the momentum transfer due to turbulence eddies can be modeled using turbulent viscosity (or eddy viscosity) $v_T$, which relates turbulent stresses to the mean flow velocities as,

$$-\langle u_i' u_j' \rangle + \frac{2}{3} k \delta_{ij} = v_T \left( \frac{\partial \langle u_i \rangle}{\partial x_j} + \frac{\partial \langle u_j \rangle}{\partial x_i} \right) \tag{6}$$

$$k \equiv \frac{1}{2} \langle u_i' u_i' \rangle \tag{7}$$

where, $\delta_{ij}$ is the Kronecker delta function and $k$ is the turbulent kinetic energy. By substituting Equation 6 & 7 in Equation 5 following equation is derived.

$$\frac{\partial \langle u_i \rangle}{\partial t} + \frac{\partial \langle u_i \rangle \langle u_j \rangle}{\partial x_j} = -\frac{1}{\rho} \frac{\partial (\langle p \rangle + \frac{2}{3} k \rho)}{\partial x_i} + (v + v_T) \frac{\partial^2 \langle u_i \rangle}{\partial x_j \partial x_j} \tag{8}$$

where, $v_{effective} = (v + v_T)$ is known as the effective viscosity. However, to solve RANS equations (Equation 4 and Equation 8), turbulent viscosity $v_T$ should be determined. Many turbulent models are available for determining turbulent viscosity and some of the most common models are described in the following section

2.1.1.1 *Standard $k - \epsilon$ model*

This model [19] is a two-equation turbulent viscosity model which includes two extra transport equations to characterize the turbulence properties. The first transport equation (Equation 9) is for turbulent kinetic energy, $k$ which determines the energy in turbulence and the second transport equation (Equation 10) is for turbulent dissipation which determines the scale of the turbulence.

$$\frac{\partial k}{\partial t} + \langle u_j \rangle \frac{\partial k}{\partial x_j} = 2v_t \overline{S_{ij} S_{ij}} - \rho \epsilon + \frac{\partial \left[ (\frac{v_t}{\sigma_k}) \frac{\partial k}{\partial x_j} \right]}{\partial x_j} \tag{9}$$

$$\frac{\partial \epsilon}{\partial t} + \langle u_j \rangle \frac{\partial \epsilon}{\partial x_j} = \frac{C_1 \epsilon}{k} 2v_t \overline{S_{ij} S_{ij}} - \frac{C_2 \rho \epsilon^2}{k} + \frac{\partial \left[ (\frac{v_t}{\sigma_\epsilon}) \frac{\partial k}{\partial x_j} \right]}{\partial x_j} \tag{10}$$

where, turbulent viscosity is defined as,

$$v_t = \frac{Ck^2}{\varepsilon} \tag{11}$$

In above equations: k is turbulent kinetic energy, $\epsilon$ is turbulent dissipation, $\overline{S_{ij}}$ is the mean strain rate tensor. The terms $C, C_1, C_2$ are closure coefficients which can be found in detail in the references [20].

### 2.1.1.2  $k - \omega$ model

This is another two-equation turbulent viscosity model suggested by Wilcox (1988) [21]. Here, two extra transport equations are modeled for turbulent kinetic energy, $k$ and specific dissipation rate, $\omega$.

$$\frac{\partial k}{\partial t} + \langle u_j \rangle \frac{\partial k}{\partial x_j} = \tau_{ij} \frac{\partial u_i}{\partial x_j} - \beta^* k \omega + \frac{\partial \left[ (v + \sigma_k v_t) \frac{\partial k}{\partial x_j} \right]}{\partial x_j} \tag{12}$$

$$\frac{\partial \omega}{\partial t} + \langle u_j \rangle \frac{\partial \omega}{\partial x_j} = \alpha \frac{\omega}{k} \tau_{ij} \frac{\partial u_i}{\partial x_j} - \beta \omega^2 + \frac{\partial \left[ (v + \sigma_\omega v_t) \frac{\partial \omega}{\partial x_j} \right]}{\partial x_j} \tag{13}$$

where, turbulent viscosity is defined as,

$$v_t = \frac{k}{\omega} \tag{14}$$

In above equations: $\tau_{ij}$ is the mean stress tensor. The terms $\alpha, \beta, \beta^*, \sigma_k, \sigma_\omega$ are closure coefficients which can be found in detail in the references [21].

### 2.1.1.3 SST $k - \omega$ model

This model is proposed by Menter (1994) [22]. This model uses $k - \omega$ formulation at the boundary layers and it switches to $k - \epsilon$ close to a free stream boundary condition as $k - \omega$ is too sensitive to free-stream boundary conditions [23]. Transport equations for $k$ and $\omega$ proposed by Menter (1994) are shown in Equation 15 and 16.

$$\frac{\partial k}{\partial t} + \langle u_j \rangle \frac{\partial k}{\partial x_j} = 2v_t \overline{S_{ij} S_{ij}} - \beta^* k\omega + \frac{\partial \left[(v + \sigma_k v_t) \frac{\partial k}{\partial x_j}\right]}{\partial x_j} \tag{15}$$

$$\frac{\partial \omega}{\partial t} + \langle u_j \rangle \frac{\partial \omega}{\partial x_j} = \alpha \overline{S_{ij} S_{ij}} - \beta \omega^2 + \frac{\partial \left[(v + \sigma_\omega v_t) \frac{\partial \omega}{\partial x_j}\right]}{\partial x_j} + 2(1 - F_1) \sigma_{\omega 2} \frac{1}{\omega} \frac{\partial k}{\partial x_i} \frac{\partial \omega}{\partial x_i} \tag{16}$$

In above equations: k is turbulent kinetic energy, $\omega$ is turbulent dissipation rate, $\overline{S_{ij}}$ is the mean strain rate tensor. The terms $\alpha, \beta, \beta^*, \sigma_k, \sigma_\omega, \sigma_{\omega 2}$ and $F_1$ are closure coefficients and auxiliary relations which can be found in detail in the references [22, 24, 25].

### 2.1.1.4 *Reynolds Stress Transport (RST) model*

This model attempts to solve the Reynolds stress term $\langle u_i' u_j' \rangle$ in RANS equations directly by calculating the Reynolds stress terms by solving the transport equation for the transport of the Reynold stresses.

$$\frac{D \langle u_i' u_j' \rangle}{Dt} = -\left( \langle u_i' u_k' \rangle \frac{\partial U_j}{\partial x_k} + \langle u_j' u_k' \rangle \frac{\partial U_i}{\partial x_k} \right) - 2\nu \frac{\partial \langle u_i' \rangle}{\partial x_k} \frac{\partial \langle u_j' \rangle}{\partial x_k} + \frac{\langle p \rangle}{\rho} \left( \frac{\partial \langle u_i' \rangle}{\partial x_j} + \frac{\partial \langle u_j' \rangle}{\partial x_i} \right) - \frac{\partial}{\partial x_k} ( \langle u_i' u_j' u_k' \rangle +$$

$$\frac{\langle p u_i' \rangle}{\rho \delta_{jk}} + \frac{\langle p u_j' \rangle}{\rho \delta_{ik}} - \nu \frac{\partial \langle u_i' u_j' \rangle}{\partial x_k} ) \tag{17}$$

However, to solve the Reynold stress transport equations, closure models are required to solve second, third and fourth terms in the right hand side of Equation 17 [26]. RST model is computationally more expensive than eddy viscosity models (such as $-\omega$, $k - \epsilon$) since it solves more equations. However, it has advantages over eddy viscosity models (like $k - \omega$) since it doesn't use an isotropic eddy viscosity and it solves for all components in turbulent transport. Hence, it accounts for turbulence anisotropy, streamline curvature, swirl rotation [27].

### 2.1.2 Large Eddy Simulation (LES)

Based on Kolmogorov theory, large scale eddies contain most of the turbulence energy. LES directly calculates the large-scale motions while smaller scales are modeled under the assumption that they behave isotopically as stated in Kolmogorov theory. To differentiate between the larger scale and smaller scale motions, LES uses a low-pass filter to decompose flow velocity as well as other flow variables.

$$\bar{u} = u - u' \tag{18}$$

In Equation 18 $\bar{u}$ represents the larger resolved scales while $u'$ is the smaller unresolved scale, which is also known as the subgrid-scale component. This is achieved by applying the following filtering operation.

$$\bar{u} = \oint u(x')G(x,x';\Delta)dx' \tag{19}$$

Where the filter function $G(x,x';\Delta)$ satisfies,

$$\oint G(x,x';\Delta)dx' = 1 \tag{20}$$

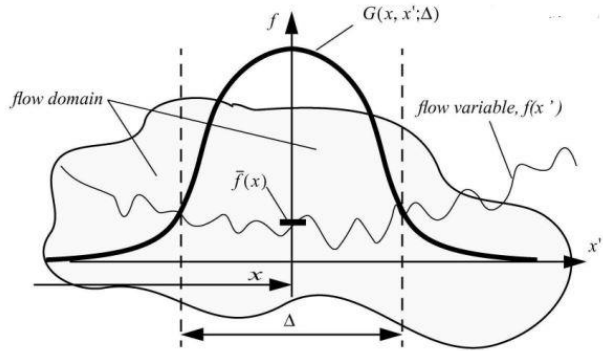

Figure 2-1: Representation of LES filter [28]

The filtered Navier-stokes equations are,

$$\frac{\partial \bar{u_i}}{\partial x_i} = 0 \tag{21}$$

$$\frac{\partial \bar{u_i}}{\partial t} + \frac{\partial \overline{u_i u_j}}{\partial x_j} = -\frac{1}{\rho}\frac{\partial \bar{p}}{\partial x_i} + \upsilon \frac{\partial^2 \bar{u_i}}{\partial x_j \partial x_j} \tag{22}$$

In Equation 22 $\bar{p}$ is the filtered pressure term while $\overline{u_i u_j}$ is a nonlinear convective term which links the resolved (larger eddies) and unresolved (smaller eddies) as following.

$$\tau_{ij}^R = \overline{u_i u_j} - \bar{u_i}\bar{u_j} \tag{23}$$

$\tau_{ij}^R$ is the sub-grid scale stress and it is decomposed as :

$$\tau_{ij}^R = \tau_{ij}^r + \frac{2}{3}k_r \delta_{ij} \qquad (24)$$

In Equation 24, the residual kinetic energy is defined as half of the trace of the sub-grid scale stress tensor, $k_r \equiv \frac{1}{2}\tau_{ii}^R$ and $\tau_{ij}^r$ is the residual stress term.

Using Equation 23 and 24 in Equation 22 can be rewritten as,

$$\frac{\partial \overline{u_i}}{\partial t} + \frac{\partial \overline{u_i u_j}}{\partial x_j} = -\frac{1}{\rho}\frac{\partial \bar{p}}{\partial x_i} + \upsilon \frac{\partial^2 \langle u_i \rangle}{\partial x_j \partial x_j} - \frac{\partial \tau_{ij}^r}{\partial x_j} \qquad (25)$$

Equation 25 is unclosed and the closure is achieved by an eddy viscosity model which relates the residual stress term $\tau_{ij}^r$ to the resolved rate of strain tensor $\overline{S_{ij}}$ with the eddy viscosity of the unresolved motions $\upsilon_r$.

$$-\tau_{ij}^r = 2\upsilon_r \overline{S_{ij}} \qquad (26)$$

$$2\overline{S_{ij}} = \left(\frac{\partial \overline{u_i}}{\partial x_j} + \frac{\partial \overline{u_j}}{\partial x_i}\right) \qquad (27)$$

Using above relations Equation 25 can be reduced to:

$$\frac{\partial \overline{u_i}}{\partial t} + \frac{\partial \overline{u_i u_j}}{\partial x_j} = -\frac{1}{\rho}\frac{\partial (\bar{p} + \frac{2}{3}k_r \rho)}{\partial x_i} + (\upsilon + \upsilon_r)\frac{\partial^2 \langle u_i \rangle}{\partial x_j \partial x_j} \qquad (28)$$

To model $\upsilon_r$, a sub-grid scale (SGS) model is needed. In this study, Smagorinsky SGS model is used to model $\upsilon_r$ since it performs well for wall bounded flows [29]. Smagorinsky model models the eddy viscosity as,

$$v_r = (C_s\Delta)^2\sqrt{2\overline{S_{iJ}S_{iJ}}} \qquad (29)$$

where, $\Delta$ is the filter length and $C_s$ is a constant, which is calibrated for isotropic turbulence [30].

2.1.3   Detached Eddy Simulation (DES)

DES is a hybrid RANS-LES computational approach which focuses on combining the advantages of both RANS and LES methods [31, 32]. Hence, DES does not require very fine mesh size as LES and computationally less expensive. Although it's computationally expensive than RANS turbulent models, its known to provide better results since ability of RANS to solve unsteady turbulent motion is limited. In DES, the model acts like RANS where the flow is in attached boundary layers and it switches to LES where flow separation is present [33]. As shown in above sections, both RANS and LES have similar formations (Equation 8 and Equation 28) and in both equations unknown eddy viscosity term ($v_T$ $or$ $v_r$) need to be modeled. These similarities in the equations allow the uniform switching between RANS and LES to solve the flow problem. The switching between the two models ( RANS model and SGS model) is done based on the local grid resolution and the distance from the wall [33]. For the current study, SST k-$\omega$ and Smagorinsky SGS model was employed in DES simulation. DES may cause some issues due to inaccurate switching to LES-mode from RANS mode inside the boundary layer. This is known to be caused due to the ambiguous grid spacing close to walls which causes the model to switch to LES mode inside the boundary layer, causing grid induced separation (GIS)[34].

## 2.2 Modelling of Flow Generated Sound

Numerical modelling of flow generated sound is often referred as Computational Aero-acoustics (CAA) modelling. Although the term "Aero-acoustics" indicates air flow, same techniques can be implemented for liquids, which may be termed hydro acoustics. CAA methods can be classified in to two groups, namely direct and hybrid methods. The direct method solves the compressible Navier-Stokes equations and together computes flow and acoustic solutions while hybrid approaches compute acoustics separately using the results from flow computation.

### 2.2.1 Direct Methods

The acoustic field in a fluid flow can be fully computed through compressible Navier-stokes equations. Such a computation will include the sound generation, propagation as well as the interaction between acoustic and flow fields. Hence, a compressible CFD simulation can be used to solve aero-acoustic problems and such methods are referred as direct CAA methods. However, direct methods have certain issues that hybrid CAA methods are often preferred over direct CAA methods. Some of these limitations are described below.

**Low time step requirement due to length scale differences:** length scales of the flow vary from the Kolmogorov length scale $l_k$ to the large eddy scale $L$. Length scales in the acoustic domain are related to the acoustical wave length $\lambda$ which is much larger than $l_k$. To accurately capture the sound sources, CFD mesh should be in the order of $l_k$ in the source regions. For the solutions to be stable Courant–Friedrichs–Lewy (CFL) number in both flow and acoustic domains should be kept to a small value (typically less than unity for explicit differencing schemes).

$$CFL_{CFD} = u.\frac{dt}{\nabla x} \qquad (30)$$

$$CFL_{CAA} = c.\frac{dt}{\nabla x} \qquad (31)$$

Equation 1 and 2 show the CFL numbers for flow and acoustic problems respectively, where $u$ is the flow velocity and $c$ is the wave velocity which is close to the sound velocity in a fluid. As, $c$ is much greater than, the time step $dt$ should be maintained at a very low value for both domains. Considering the grid size $\nabla x$ is same for both domains. Hence, excessive computational power is needed for the simulations.

**Energy differences:** The energy levels in the acoustic field are much smaller compared to the flow field. For low Mach number (M) flow, this difference increases since the acoustic power is in the order of $M^3$ and $M^4$ [35]. When the acoustic variables are too small, numerical errors in the simulation can interfere with acoustic results. Hence, numerical error should be maintained much smaller than the acoustic variables, which takes extra computational effort.

**Boundary conditions**: The boundary conditions for CFD simulation generates spurious numerical reflections in the acoustic domain [35], since both flow and acoustics are solved in the same simulation. Hence, special modifications are needed for CFD boundary conditions, to reduce these spurious effects on wave propagation.

2.2.2 Hybrid CAA Methods

The following section describes several hybrid CAA methods, starting from the theoretical description of acoustic wave propagation in a fluid.

Wave equation describes the propagation of wave in a uniform stagnant flow field. Wave equation can be used to describe the behavior of waves occur in different physics such as sound waves, light waves, water waves. This wave propagation can be observed in fields such as electro magnetics, fluid dynamics and acoustics.

$$\left(\frac{1}{c_o^2}\frac{\partial^2}{\partial t^2} - \nabla^2\right)\varphi' = 0 \tag{32}$$

Equation 32 is the homogeneous wave equation with a fluctuation variable $\varphi'$. In a fluid, that variable either can be pressure, velocity or density. $c_0$ is the sound velocity of the fluid. When the right-hand side of the acoustic wave equation is not zero, it is called non-homogeneous wave equation (Equation 33). Terms in the right-hand side $f(\vec{x}, t)$ represent the acoustic sources.

$$\left(\frac{1}{c_o^2}\frac{\partial^2}{\partial t^2} - \nabla^2\right)\varphi' = f(\vec{x}, t) \tag{33}$$

For a fluid with stagnant mean flow (mean velocity is zero) the wave equation can be derived for small perturbations of the flow. The linearized equations for mass and momentum continuity as well as the energy equation for small perturbations of flow quantities can be written as in equations 34,35,36.

$$\frac{\partial \rho'}{\partial t} = -\rho_o \frac{\partial v_i'}{\partial x_i} \tag{34}$$

$$\rho_o \frac{\partial v_i'}{\partial x_i} = -\frac{\partial v_i'}{\partial x_i} + f_i \tag{35}$$

$$\frac{\partial s'}{\partial t} = \frac{Q_w}{\rho_o T_o} \tag{36}$$

In above equations, $\rho', v', s'$ denote density, velocity and entropy perturbations. $f_i, Q_w$ denote external force and heat source terms respectively. $\rho_o, T_o$ denote the density and temperature at reference state.

The linearized equation of state can be written as follows, where $p'$ is the pressure perturbation.

$$p' = c_o^2 \rho' + \left(\frac{\partial p}{\partial s}\right)_\rho s' \tag{37}$$

Differentiating Equation 34 w.r.t "t" and subtracting the divergence of Equation 35 and then substituting for $\rho'$ using Equation 37, following equation can be derived.

$$\frac{1}{c_o^2}\frac{\partial^2 p'}{\partial t^2} - \frac{\partial^2 p'}{\partial x_i^2} = \frac{1}{c_o^2}\left(\frac{\partial p}{\partial s}\right)_\rho \frac{\partial^2 s'}{\partial t^2} - \frac{\partial f_i}{\partial x_i} \tag{38}$$

Furthermore, using the relation from Equation 37, Equation 38 can be further reduced to Equation 39.

$$\frac{1}{c_o^2}\frac{\partial^2 p'}{\partial t^2} - \frac{\partial^2 p'}{\partial x_i^2} = \frac{1}{\rho_o c_o^2 T_o}\left(\frac{\partial p}{\partial s}\right)_\rho \frac{\partial Q_w}{\partial t} - \frac{\partial f_i}{\partial x_i} \tag{39}$$

Equation 39 implies that the external force field terms as well as time varying heat production terms act as sound sources. In absence of these terms the homogeneous wave equation can be obtained.

$$\frac{1}{c_o^2}\frac{\partial^2 p'}{\partial t^2} - \frac{\partial^2 p'}{\partial x_i^2} = 0 \tag{40}$$

### 2.2.2.1 *Lighthill's acoustic analogy*

The first hybrid CAA method was introduced by sir James Lighthill in 1952 [36]. This method is referred as Lighthill's acoustic analogy. The term "analogy" is used since the acoustic field is described using classical wave equations, which describe propagation of perturbations compared to a reference state. Lighthill rearranged Navier Stokes equations (in absence of external forces and heat sources) using vector operations and derived a similar equation to the non-homogeneous wave Equation. The terms in the right-hand side is identified as the sound sources.

$$\frac{\partial^2 \rho'}{\partial t^2} - c_o^2 \frac{\partial^2 \rho'}{\partial x_i^2} = \frac{\partial^2 T_{ij}}{\partial x_i x_j} \tag{41}$$

Where density perturbation is defined as,

$$\rho' = \rho - \rho_o \tag{42}$$

Where $\rho$ is the instantaneous density and $\rho_o$ is the reference density.

$$T_{ij} = \rho U_i U_j + [\, p' - c_o^2 \rho'\,]\delta_{ij} - \tau_{ij} \tag{43}$$

$T_{ij}$ is called the Lighthill's stress tensor, which represents the acoustics sources generated by the turbulence, which are described as quadrupole type volumetric sound sources. In Equation 43, subscripts *i,j* denotes the Cartesian vector components while $U$ denotes the convective velocity and $p'$ denotes the pressure perturbations (which is defined similar to Equation 42). The viscous stress tensor $\tau_{ij}$ for a Newtonian fluid is defines as,

$$\tau_{ij} = \rho v \left( \frac{\partial U_i}{\partial x_j} + \frac{\partial U_j}{\partial x_i} - \frac{2}{3}\left(\frac{\partial U_k}{\partial x_k}\right)\delta_{ij} \right) \tag{44}$$

The first term of the Lighthill stress tensor $\rho U_i U_j$ is very similar to Reynold's stress terms. The second term represents the entropy fluctuations $s' = p' - c_o^2 \rho'$ and the third term is the viscous shear stresses caused by the velocity gradients.

As the right-hand side and left-hand side in Equation 41 spatially separate the propagation part and the source part of the acoustic domain, respectively, the following assumptions/limitations can be inferred [37].

The term , $\rho U_i U_j$ in the source part consists the convective velocity terms. Hence, it is assumed that the convective velocity in the propagation part is zero.

- The term, $s' = p' - c_o^2 \rho'$ in the source part infers that entropy loss in propagating part is negligible or zero.
- The term, $\tau_{ij}$ in the source part infers that there are no viscous losses in the propagating part.
- Lighthill equations are derived assuming the absence of external forces and heat sources.

### 2.2.2.2 *The Ffowcus Williams-Hawking's analogy*

FW-H analogy is an extension of Lighthill's analogy where additional source terms are considered when the flow interacts with solid surfaces (moving or stationary) [38, 39]. FW-H formulation is shown in Equation 9 and detailed derivation of this formula can be found in [40].

$$\frac{\partial^2 \rho'}{\partial t^2} - c_o^2 \frac{\partial^2 \rho'}{\partial x_i^2} = \frac{\partial^2 [T_{ij} H(f)]}{\partial x_i x_j} - \frac{\partial}{\partial x_i}\left(p' - \delta_{ij}\tau_{ij}\right)\delta(f)\frac{\partial f}{\partial x_j} + \frac{\partial}{\partial t}\left(\rho_0 v_i \delta(f)\frac{\partial f}{\partial x_i}\right) \quad (45)$$

The function $f(x, t) = 0$ represents a permeable surface moving with velocity v, which encloses the noise sources on solid surfaces (see Figure. 2-2). While f = 0 defines the control surface, $f > 0$ defines the domain outside the control surface and $f < 0$ defines the domain inside the control surface. $H(f)$ is the Heaviside function, which is zero inside the control surface and unity elsewhere. $\delta(f)$ represents the kroneker delta function which is unity on the surface f=0 and zero elsewhere. $v_i$ is the surface velocity. The domain description is presented in Figure .2-2.

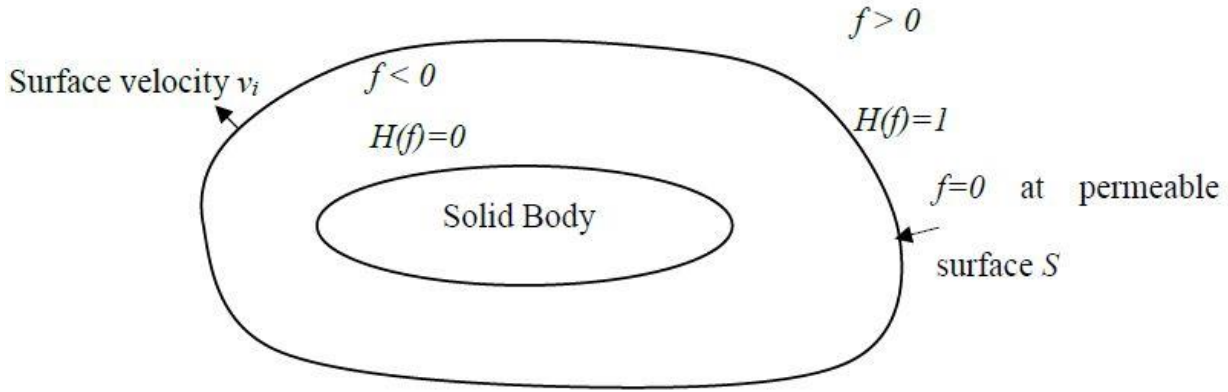

Figure 2-2 : FW-H domain representation

The first source term in the right-hand side is the same source term presented in Lighthill's analogy. In FW-H analogy, this source term is only considered outside the permeable surface S ,where $H(f)=1$. The second and third source terms describe the sound generated from fluid forces on the surface and mass flow fluctuations on the surface, respectively. The control surface either can be selected as a permeable surface such that it encloses a solid surface or on the solid surface. According to the above equation, when the control surface coincides with a solid surface, mathematically it can be interpreted that the monopole type sources are generated due to the volume displacement of the solid body and the dipole sources are due to the pressure fluctuations on the solid surface.

The solution fo FW-H equation is as follows.

$$p'(x,t) = \frac{1}{4\pi} \frac{\partial^2}{\partial x_i x_j} \iiint_V \left[\frac{T_{ij}}{r(1-M_r)}\right]_{t_e} dV - \frac{1}{4\pi} \frac{\partial}{\partial x_i} \iint_S \left[\frac{p'-\delta_{ij}\tau_{ij}}{r(1-M_r)}\right]_{t_e} ds + \frac{\partial}{\partial t} \frac{1}{4\pi} \iint_S \left[\frac{\rho_0 v_i \frac{\partial f}{\partial x_i}}{r(1-M_r)}\right]_{t_e} ds \quad (46)$$

In above equation, $r = |x - y|$, where $x$ and $y$ denote the locations of the receiver and sound source, respectively. In the first term, the integral is taken over a volume $V$, outside the surface $S$ where $H(f)=1$. Volume V can be selected such that it encloses all Lighthill source terms. The intergals of the second and third terms refers to the intergals taken over the surface $S$ where $f=0$. $t_e$ represents the retarded time $t_e = t - \frac{r}{c}$, which is the minus the time taken for a wave to propagate from the source to the receiver. In Equation 10, $M_r = \frac{v_r}{c}$, where $v_r$ is the relative velocity of the receiver with respect to the source. The solution presented in Equation 46 is widely referred as the intergal solution of the FW-H. This method allows to decompose the calculated acoustic pressure, based on the transmitted pressures from each source terms. For, example, if the focus is on the Lighthill sources, intergals for other source terms can be ommited.

Since, FW-H is an extended analogy of Lighthill's analogy, all the assumptions made for Lighthill's analogy are unchanged for FW-H analogy. In addition, the intergal solution limits the usage of this method for confined flow problems, where the sources are bounded ?by solid surfaces. This is because, the solution assumes that there are no obstacles between surface S, volume $V$ and the receiver location .

### 2.2.2.3 *Acoustic Perturbation Equations (APE)*

A set of Acoustic Perturbation equations (APEs) were introduced by R. Ewert and W. Schroder (2004) [41] to simulate acoustic fields in space and time where the acoustic sources are to be calculated from an unsteady flow simulation. In their formulations they incorporated a flow decomposition, where the velocity perturbations are split into an irrotational and solenoidal part based on Helmholtz decomposition theorem which states that a vector field in 3D can be decomposed into sum of irrotational ($\nabla \times u^a = 0$) and solenoidal ($\nabla . u^v = 0$) parts. The irrotational part $u^a$ of the fluctuation is completely related to the acoustical mode and the solenoidal part $u^v$ is related to aerodynamic or turbulent fluctuating part [41].

$$u = \bar{u} + u' = \bar{u} + u^v + u^a \qquad (47)$$

Where, $\bar{u}, u', u^v, u^a$ denote time averaged mean, total perturbation, solenoidal vortical perturbation and irrotational acoustic perturbation of velocity, respectively. Then the continuity equation can be written as,

$$\frac{\partial \rho'}{\partial t} + \nabla . (\bar{\rho} u^a + \rho' \bar{u}) = -\nabla . (\bar{\rho} u^v) \qquad (48)$$

Based on a source filtering technique R. Ewert and W. Schroder [41] formulated a system of APEs for the for the variables $p'$ and $u^a$.

$$\frac{\partial p'}{\partial t} + c^2 \nabla \left( \bar{\rho} u^a + \bar{u} \frac{p'}{c^2} \right) = c^2 q_c \qquad (49)$$

$$\frac{\partial u^a}{\partial t} + \nabla(\bar{u} u^a) + \nabla \left( \frac{\rho'}{\bar{\rho}} \right) = q_m \qquad (50)$$

where the source terms,

$$q_c = \underbrace{-\nabla\rho.u^v}_{I} + \underbrace{\frac{\gamma\bar{p}}{c_p}\frac{Ds'}{Dt}}_{II} \tag{51}$$

$$q_m = \underbrace{\nabla\phi_p}_{III} + \underbrace{(-\bar{\omega}\times u^a)}_{IV} + \underbrace{T'\nabla\bar{s} - s'\nabla\bar{T}}_{V} \tag{52}$$

$$\nabla^2\phi_p = -\nabla\left[(\bar{u}.\nabla)u^v + +(u^v.\nabla)\bar{u} + \left((u^v.\nabla)u^v\right)' - (\frac{\nabla.\tau}{\rho})'\right] \tag{53}$$

In the above equations: $\rho,p,c,\ s,T,\ \omega,c_p,\gamma,\tau$ denote density, pressure, sound velocity, entropy, temperature, vorticity, specific heat constant, heat capacity ratio and shear stress, respectively.

Using the above system of equations, different types of APEs were derived [41] for different types of flow simulations (compressible, incompressible, combustion, etc.), where the acoustic sources are to be determined using CFD. Given the scope of the present work, only the APEs derived for use in incompressible unsteady simulation for low Mach number flow are discussed here. R. Ewert and W. Schroder presented this as APE variant-2 system where the perturbation pressure is decomposed as following.

$$p' = \bar{\rho}\phi_p + p^a = P' + p^a \tag{54}$$

Where, $\bar{\rho}\phi_p = P'$ is the incompressible pressure perturbation also identified as hydrodynamic pressure perturbation. $P'$ is also called "pseudo sound" and will be discussed further in section 2.2.3. In Equation 54, $p^a$ is the acoustic perturbation pressure excluding the pseudo sound component.

For incompressible flow without heat sources all the source terms are ignored except for source term *III* [41]. Under these assumptions, Equation 49 and 50 can be rewritten as,

$$\frac{\partial p^a}{\partial t} - c^2 \nabla \left( \bar{\rho} u^a + \bar{u} \frac{p'}{c^2} \right) = 0 \tag{55}$$

$$\frac{\partial u^a}{\partial t} + \nabla(\bar{u} u^a) + \nabla \left( \frac{\rho'}{\bar{\rho}} \right) = \frac{\nabla P'}{\bar{\rho}} \tag{56}$$

Taking the total derivative $\frac{D}{Dt} = \left( \frac{\partial}{\partial t} + \bar{u} . \nabla \right)$ of Equation 55 and divergence $\nabla$ of Equation 56, two equations can be obtained with terms $p^a$ and $u^a$. Then these equations can be simplified for low Mach number flow using the incompressibility assumption $\nabla . \bar{u} = 0$. Using these simplified equations, the following equation can be derived for the acoustic pressure, $p^a$.

$$\frac{1}{c^2} \frac{\partial^2 p^a}{\partial t^2} + \frac{2(\bar{u}.\nabla)}{c^2} \frac{\partial p^a}{\partial t} + \frac{\bar{u}.\nabla}{c^2} (\nabla . \bar{u} p^a) - \nabla^2 p^a = \frac{1}{c^2} \frac{\partial^2 P'}{\partial t^2} - \frac{2(\bar{u}.\nabla)}{c^2} \frac{\partial P'}{\partial t} + \frac{\bar{u}.\nabla}{c^2} (\nabla . \bar{u} P') \tag{57}$$

Unlike Lighthill's analogy, APE considers the convection and refraction effects of a non-uniform flow field [41] as the convection effects are encoded in the products of mean flow and acoustic perturbations in the right-hand side of Equation 57.

For very low Mach number flows $\bar{u} \ll c$, the convective effects can be considered negligible and Equation 57 can be reduce to,

$$\frac{1}{c^2} \frac{\partial^2 p^a}{\partial t^2} - \nabla^2 (p^a) = -\frac{1}{c^2} \frac{\partial^2 (P')}{\partial t^2} \tag{58}$$

The source term in the right-hand side is the time derivative of the incompressible pressure perturbation, in contrast to the Lighthill's source term. $P'$ is the incompressible pressure perturbation that is identified as hydrodynamic pressure perturbation and also called "pseudo sound".

2.2.3  Pseudo Sound

The sources of the flow generated sound are often localized to a region in the flow domain and only part of the energy associated with flow fluctuation radiates at the speed of sound. In the subject of flow generated sound, a terminology is used to discern propagating and non-propagating pressure fields as sound and pseudo-sound or acoustic pressure and hydrodynamic pressure [38, 42, 43].  In other words, the acoustic pressure is the pressure fluctuations that propagate away from the source region with the sound speed. In contradistinction, the pseudo-sound or hydrodynamic pressure does not propagate with the sound speed. Instead that pressure travel at the speed of the eddies in the source region. Pseudo-sound only exists in the sound source region and due to the pressure fluctuations balancing the local fluid accelerations. However, in the near-field a microphone can detect both pseudo sound and the propagating sound [38, 42].  Pseudo-sound is associated with hydrodynamic pressure fluctuations, since the term hydrodynamic is a reminder the dynamics of the generated noise is rarely influence by the compressibility[42, 43]. Another way to distinguish between acoustic and hydrodynamic fluctuations is that the acoustic pressure does satisfy the linear wave equation and hydrodynamic pressure fluctuations does not [44]. In the study by Fredburg [10], it is mentioned that the near-field pressure (or the pressure from an incompressible solution) cannot propagate since the velocity and pressure are always out of phase. While many studies have focused on decomposing acoustic and hydrodynamic pressure based on various signal processing methods such as wavelet based filtering [44, 45], theoretical decomposition of acoustic and hydrodynamic pressure can be found in early studies based on fundamental theories of flow generated sound.

In 1962, Ribner [43] proposed a way to mathematically distinguish between the acoustic and hydrodynamic pressure fluctuations in flow generated sound, starting from the Lighthill's equation (including force and mas source terms) as shown in Equation 59, where the first term in the right hand side indicates the turbulent sound sources in the flow while the second and third terms indicate the sources due to body forces in solid boundaries and sources by fluctuating mass respectively.

$$\frac{\partial^2 \rho'}{\partial t^2} - \frac{\partial^2 p'}{\partial x_i^2} = \frac{\partial^2 T_{ij}}{\partial x_i x_j} - \frac{\partial F_i}{\partial x_i} + \frac{\partial m}{\partial t} \tag{59}$$

The pressure fluctuations are distinguished in to acoustic pressure and hydrodynamic pressure.

$$p' = p^a + P^h \tag{60}$$

And the pseudo sound was defined as,

$$-\frac{\partial^2 P^h}{\partial x_i^2} = \frac{\partial^2 T_{ij}}{\partial x_i x_j} - \frac{\partial F_i}{\partial x_i} + \frac{\partial m}{\partial t} \tag{61}$$

where, $P^h = P - \bar{P}$ , calculated from an incompressible flow solution and defining $\rho' = \rho^a + \rho^h$.

which follows,

$$\frac{\partial^2 \rho'}{\partial t^2} - \frac{\partial^2 p'}{\partial x_i^2} = -\frac{\partial^2 P^h}{\partial x_i^2} \tag{62}$$

and

$$\frac{\partial^2 \rho^a}{\partial t^2} - \frac{\partial^2 P^a}{\partial x_i^2} = -\frac{\partial^2 \rho^h}{\partial t^2} \tag{63}$$

where, $\rho^a, \rho^h$ are associated with $P^a, P^h$ respectively. By using the relationship of equation of state with no heat sources following equations can be derived.

$$\frac{\partial^2 \rho^a}{\partial t^2} = \frac{1}{c^2}\frac{\partial^2 P^a}{\partial t^2} \tag{64}$$

$$\frac{\partial^2 \rho^h}{\partial t^2} = \frac{1}{c^2}\frac{\partial^2 P^h}{\partial t^2} \tag{65}$$

Substituting in Equation (63) following can be derived.

$$\frac{\partial^2 \rho^a}{\partial t^2} - \frac{\partial^2 P^a}{\partial x_i^2} = -\frac{1}{c^2}\frac{\partial^2 P^h}{\partial t^2} \tag{66}$$

$$\frac{1}{c^2}\frac{\partial^2 P^a}{\partial t^2} - \frac{\partial^2 P^a}{\partial x_i^2} = -\frac{1}{c^2}\frac{\partial^2 P^h}{\partial t^2} \tag{67}$$

This implies the radiated acoustic pressure filed is influenced by spatial distribution of source strength $-\frac{\partial^2 \rho^h}{\partial t^2}$ per unit volume. $\rho^h$ is described as the "zero-order density" perturbation in the fluid, which is related to the pseudo sound pressure $P^h$ through the equation of state, where $P^h$ to be calculated by incompressible type flow Equation. In conclusion, Equation (67) indicates that acoustic pressure radiation flow generated sound sources originate from the source term $-\frac{1}{c^2}\frac{\partial^2 P^h}{\partial t^2}$ and it is irrelevant how the sound is produced, whether due to turbulence flow effects, body forces in the boundary or mass sources in the flow [43].

## 2.3 Flow Decomposition

### 2.3.1 Proper Orthogonal Decomposition

Proper Orthogonal Decomposition (**POD**) can be used as a tool to analyze the sound sources by decomposing appropriate fluid flow quantities in to spatial modes with respective time evolvements. POD has been used to analyze sound sources in fluid flow in previous work [46, 47]. Following is summary of the procedure of applying POD for a numerical simulation.

$$u(x,t) \cong \{\sum_{i=0}^{M-1} \mu_i(t) \emptyset_i(x)\} \tag{68}$$

The quantity $u(x,t)$ can be presented as a summation of orthogonal (linearly independent) mode shapes $\emptyset_i(x)$ multiplied by their time varying amplitudes $\mu_i(t)$, such that the mode shapes will perfectly represent $u(x,t)$, when M $\to \infty$. POD is focused on finding the best possible highest energy containing M orthogonal modes to represent $u(x,t)$. The modes $\emptyset_i(x)$ represents the coherent fluid structures of the fluid, within the time duration that POD is calculated.

If the values for $u(x,t)$ are saved in the snapshot matrix, $0^{th}$ mode denotes the mean part (with zero frequency) and the rest denotes the fluctuating structures in the flow. If the values for $u'(x,t)$ are saved in the snapshot matrix $0^{th}$ mode will also represent a fluctuating coherent flow structure. Following is the procedure used for calculating POD modes.

1. Create a snapshot matrix $U_m$

$U_m$: First column is a flow parameter, $u$, (e.g. pressure) at all points, next column is the same but at a next time step (*dt* or multiples of *dt*). $U_m$ contains M time steps (columns).

2. Find V: Perform singular value decomposition (SVD) on snapshot matrix $U_m$

$$U_m = VDW^T \qquad (69)$$

Matrix V and W are orthonormal matrices and D is a diagonal matrix which contains the singular values. If $U_m$ is a [nxM] matrix, ideally matrix, the dimensions of V,D,W are [nxn] ,[nxM], [MxM] respectively.

3. Calculating POD modes

Columns of matrix V are the POD modes $\emptyset_i(x)$ of the system arranged from the highest energy mode to the lowest energy modes in descending order. The terms in diagonal matrix D contains the energy percentage of each mode. Since V [nxn] is a very large matrix, usually it's limited to [nxM] for calculation purposes. This will allow only to consider highest energy containing M modes.

$$V = \emptyset = [\emptyset_0(x) \ldots\ldots\ldots\ldots\ldots \emptyset_{M-1}(x)] \qquad (70)$$

4. Calculating time coefficients

In POD the time coefficients of each mode can be found as following using the orthogonality condition (i.e: transpose of the matrix is equal to the inverse) of the mode matrix.

$$[\mu_i(t_0) \quad \mu_i(t_1) \quad \ldots\ldots\ldots\ldots \quad \mu_i(t_m)]\, \emptyset_i = U_m \qquad (71)$$

$$[\mu_i(t_0) \quad \mu_i(t_1) \quad \ldots\ldots\ldots\ldots \quad \mu_i(t_m)] = \emptyset_i^T U_m \qquad (72)$$

The time coefficients of the POD modes describe the evolution of each mode with time which can be used to analyze the modes in spectral domain. The dimension of the matrix which contains the time coefficient is [MxM]. Since V is an orthonormal matrix, both the amplitude and the time evolution of modes are represented from the time coefficients.

# CHAPTER 3: EXPERIMENT AND SIMULATION PROCEDURE

## 3.1 Experimental Procedure

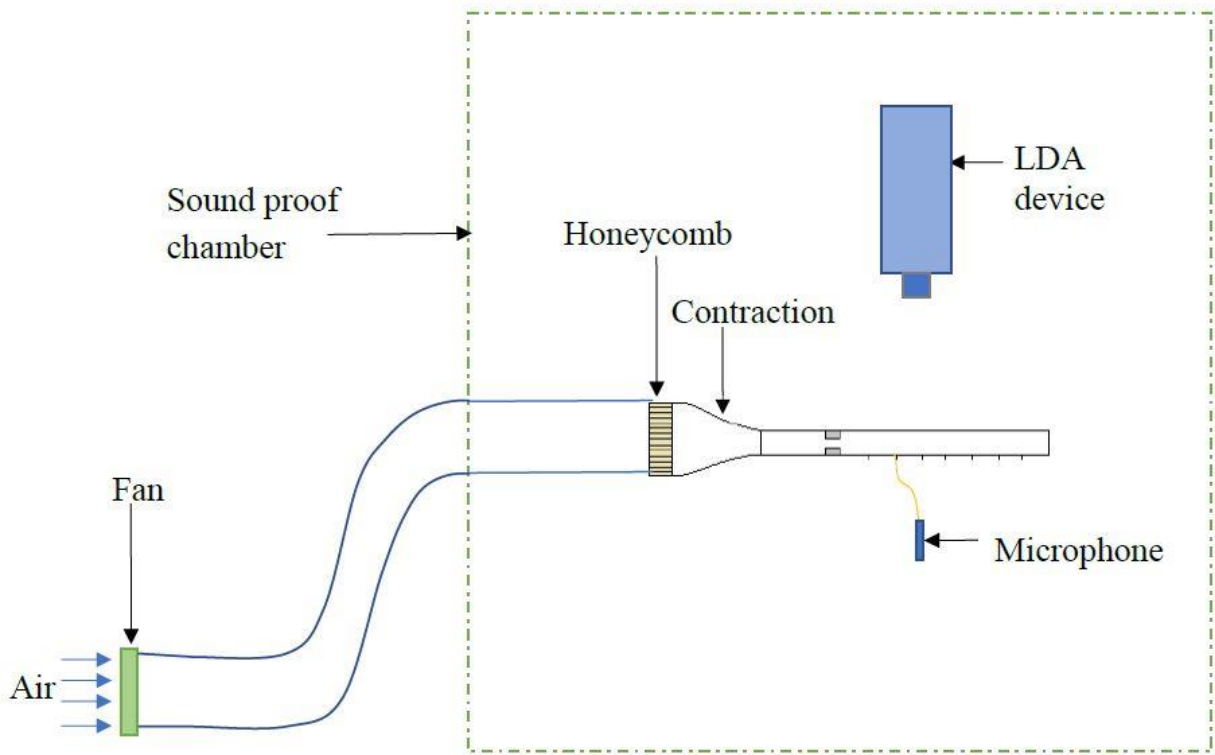

Figure 3-1: Schematic of the experimental setup

A schematic of the experimental setup is shown in Figure 3-1. Air flow was supplied by using a fan (Model: UF12A12-BTL). The test section was placed inside a sound isolation chamber (WhishperRoom$^{TM}$, model 4872). Airflow was guided through a ventilation silencing system (WhishperRoom$^{TM}$, model 4230) allowing a smooth airflow while reducing the fan noise carried with the flow. Before proceeding experiments, sound pressure was measured inside the sound proof chamber (at the outlet of the contraction) and it was confirmed that sound pressure level of the fan noise is negligible and won't interfere with the experimental measurements.

3.1.1 Velocity Measurements

In the current study Laser Doppler Anemometry (LDA) was used to measure the velocity since it's a non-intrusive method which does not interfere with the flow field and sound generation. Figure. 3-2, shows the configuration of a LDA unit.

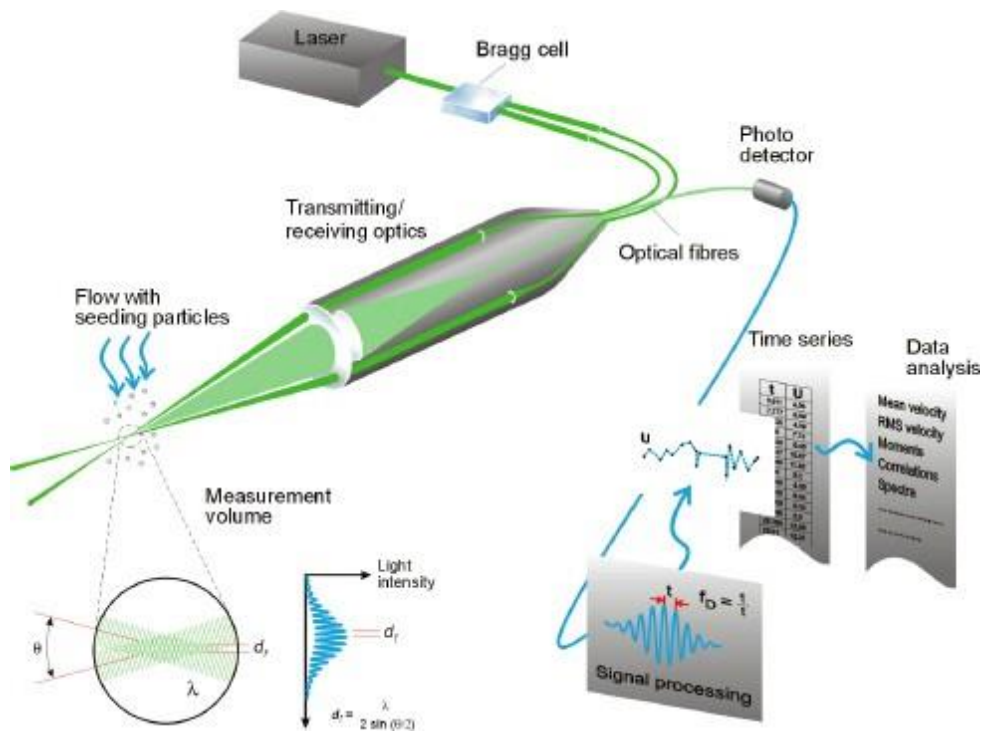

Figure 3-2: LDA system [48]

A "Bragg cell" is used to split the laser beam in to two beams with same intensity, but with a frequency shift. These, two parallel beams are focused by a lens to form a probe volume which is few millimeters long where the beams intersect. The beam intersection generates parallel planes with high light intensity which are called "fringes". Distance between fringes can be estimated based on the wave length and the angles between the beams. When seeding particles pass through this region they scatter the light, which create a doppler shift where the doppler frequency is proportional to the particle velocities. A photodetector collects the scattered light and converts the light intensity to an electrical signal. The noise from other wave lengths such as ambient light is filtered prior to photo detector. The output electrical signal from the photo detector is called the "doppler burst" signal which is later processed to determine the doppler frequency shift of each seeding particle crossing the probe volume. Based on the fringe distance and the doppler frequency shift velocity of each seeding particle is calculated.

In the current study one dimensional LDA system (Dantec Dynamics A/S., Skovulunde, Denmark) was used. The output of the Bragg cell formed beams with a wave length of 600 nm and a frequency shift of 80 MHz. Two-component fiber optic transceiver (Model FlowExplorer; Dantec Dynamics A/S., Skovulunde, Denmark) with a 300-mm focal length lens produced a probe volume with minor and major axes of 0.1 mm and 1 mm. Velocity was measured in a clear glass pipe with a 2mm thickness. A fog generator was used to generate seeding particles.

3.1.2 Sound Measurements

A ER-7C probe mic system (Etimotic research, inc) was used to measure sound. The glass tube wall was drilled and pressure port connections were inserted to connect the microphone to acquire the sound measurements at inner wall surface. Data were acquired using a NI9215 DAQ at a 15 kHz sampling frequency. Microphone was calibrated against a Larson Davis model 831 sound pressure meter. The calibration plot showed that the output of the microphone measurements are approximately 30 dB less than the actual sound pressure level for frequencies higher than 50 Hz and the sensitivity of the microphone exponentially reduced for frequencies lower than 50 Hz. The calibration data are shown in Figure. 3-3.

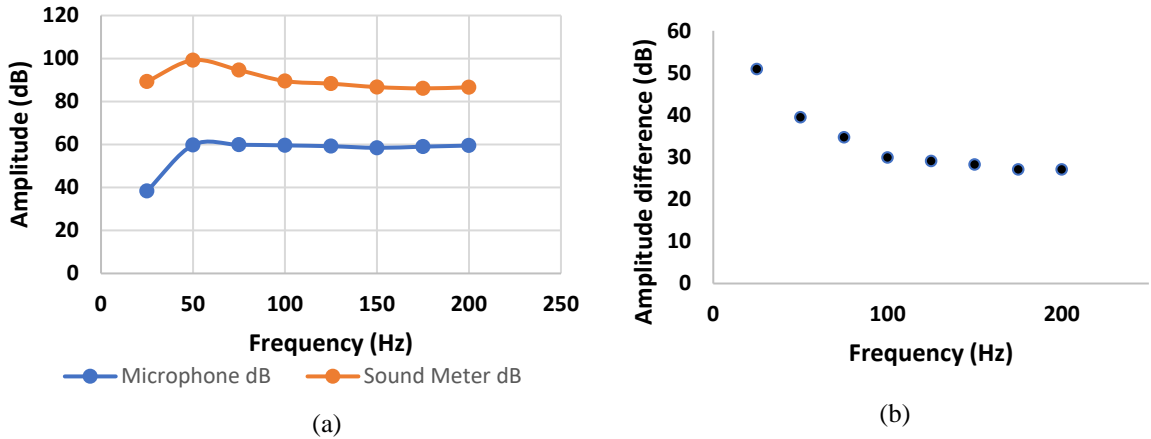

Figure 3-3: Microphone calibration (a) sound pressure level amplitudes of microphone and sound pressure meter (b) sound pressure level difference between microphone and sound pressure meter

## 3.2 Simulation Procedure

### 3.2.1 Model Geometry

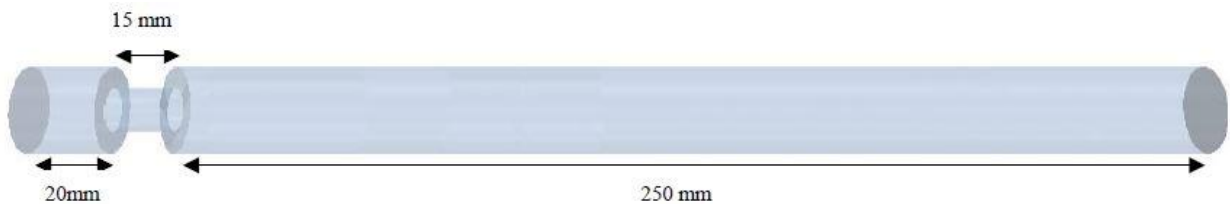

Figure 3-4: Model Geometry

Figure. 3-4 shows the geometry of the fluid domain considered in the simulation. This simple geometry was selected to represent physiological conditions such as stenosed arteries and airway stenosis. Similar geometries have been used in previous studies to investigate such physiological conditions [8]. The diameter of the tube (unconstructed part) was selected as 20.6 mm, which is comparable to the diameter of human trachea. The diameter of constricted section is 10.3 mm, which forms an area reduction of 75%.

3.2.2 Meshing

The meshing tool in StarCCM+ was employed. The geometry was meshed using polyhedral type cells. For mesh generation, arbitrary polyhedral type cells are used to build the core mesh where typically polyhedral cells have an average of 14 faces. Since polyhedral cells are bounded by many neighboring cells, the approximation of gradients in the flow is much better than other cell types such as tetrahedral [49]. Polyhedral cells are also known to perform well for complex swirling flows [49].

3.2.2.1 *Grid independence study*

Grid independence study was conducted using 4 meshes which are made progressively finer to each other (Mesh 1~ 0.6 Million, Mesh 2~ 1 Million, Mesh 3~ 1.6 Million, Mesh 4~ 2 Million cells). To evaluate the meshes, axial velocities on a line probe normal to the flow direction at 9 cm downstream the constriction and axial velocities downstream the constriction on the centerline of the duct were plotted. First, the grid independence study was carried out using steady RANS SST k-ω model and then for LES simulation. For the LES simulation mean velocity results were evaluated after the solution time reached 1 s. The results are shown in Figure 3-4 to 3-7.

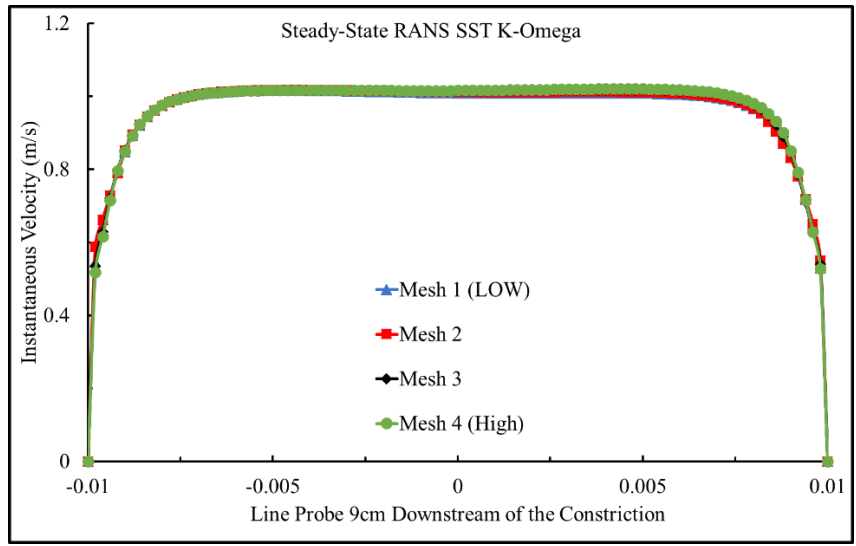

Figure 3-5 : Grid independence study: steady SST k-ω: axial velocity profile 9cm downstream the constriction

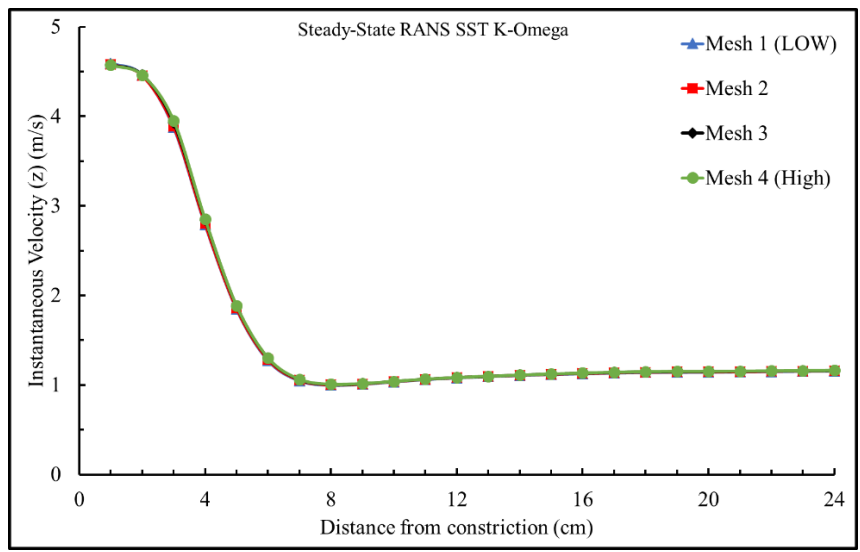

Figure 3-6: Grid independence study: steady SST k-ω: axial velocity plotted on the centerline downstream the constriction

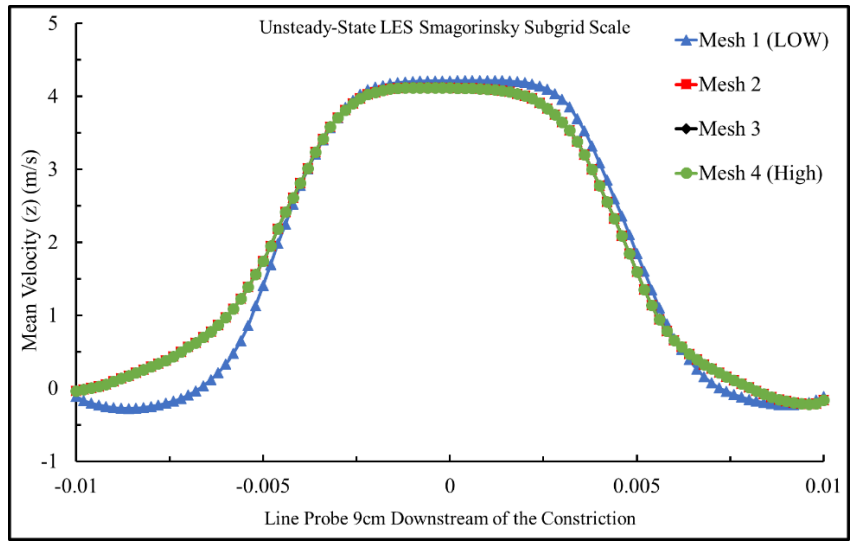

Figure 3-7: Grid independence study: LES: mean axial velocity profile 9cm downstream the constriction

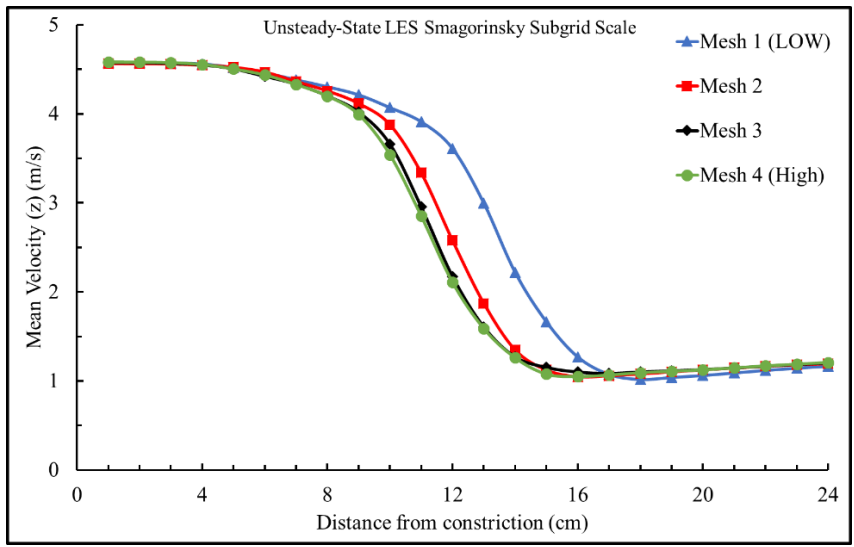

Figure 3-8: Grid independence study: LES: mean axial velocity plotted on the centerline downstream the constriction

After the grid convergence study, Mesh 4 with ~2 Million cells was selected for simulations. Here, a mesh refined region was created at the constriction and downstream the construction where the flow instabilities are expected. The average mesh size in the refined region was 0.3 mm. As the Re number at the constriction is 2340, an estimation for smallest eddy scale $\eta$ was calculated to be 0.0306 mm knowing the largest eddy scale $l$ is the diameter of the tube and using the relation $\sim l/Re^{\frac{3}{4}}$. Hence, the refined mesh size is in the order of inertial range length scale which is suitable for LES simulation. A cross-section of the meshed geometry is shown in Figure. 3-4.

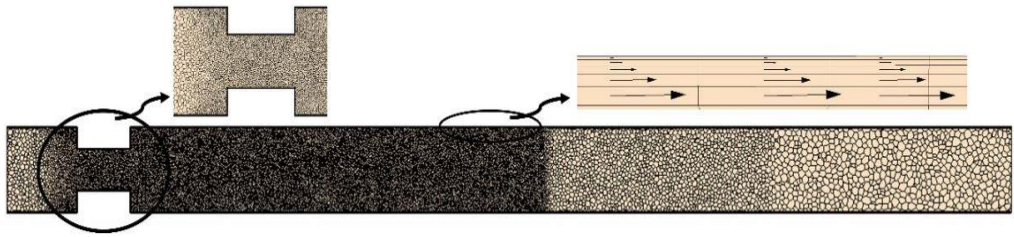

Figure 3-9: A cross section of the CFD mesh with zoomed views near the constriction and wall boundary

As shown in the Figure 3-4, a prism layer (boundary layer) mesh was added at the walls as it's important for resolving turbulent boundary layer accurately [50]. A 5 layer, prism layer mesh was employed with a total thickness of 0.3mm with an initial layer thickness of 0.025 mm. The layer thickness was progressively increased with a stretching factor of 1.5 to resolve the velocity gradients smoothly. Y+ value was maintained in the order of 1.

### 3.2.3 Computational Fluid Dynamics (CFD) Simulation

Four different turbulent models were employed in CFD software package StarCCM+ to simulate the flow field.

- RANS SST k-ω
- RANS RST
- DES SST k-ω
- LES

Governing Equations for each turbulence model can be found in chapter 2. For all equations second order temporal and special discretization was used. The density and the dynamic viscosity of the air was set to 1.184 kgm$^{-3}$ and 1.855 E-5 Pa.s, respectively.

#### 3.2.3.1 *Flow boundary conditions*

A velocity inlet boundary condition was used at the inlet. This velocity value was selected based on the measured flow rate at the outlet. Flow rate was calculated by measuring the velocity profile at the outlet using LDA measurements. Figure. 3-5 shows the measured velocity profile at the outlet.

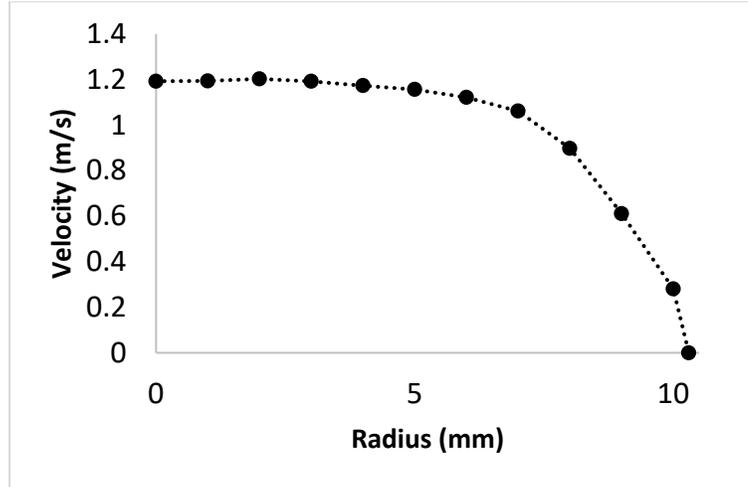

Figure 3-10: Measured outlet velocity profile

Using the outlet velocity measurements, the outlet velocity profile $U(r)$ was estimated as a function of radius $r$,

$$U(r) = -0.0004r^4 + 0.0047r^3 - 0.0215r^2 + 0.0319r + 1.1889$$

Then the mean velocity at the inlet was estimated as,

$$U_{avg} = \frac{2}{R^2}\int_0^R U(r)\, r\, dr = 0.89 \text{ ms}^{-1}$$

This inlet velocity corresponds to an inlet Reynolds number of 1170 and a Reynolds number of 2340 at the constriction. Based on previous experimental study by Saad [51], turbulence is expected for a similar geometry for inlet Reynolds numbers above 1000. Zero static pressure and non-slip boundary conditions were employed at the outlet and wall, respectively. A time step of 0.0001 was employed. For RANS simulations, turbulent intensity at the inlet was selected as 0.05 based on LDA measurements.

### 3.2.1 Computational Aero-acoustics (CAA) Simulation

CAA simulation was performed in parallel with LES using Acoustic Perturbation Equation (APE) (Equation 58) based hybrid CAA method. CAA solver was started after 5 seconds allowing the flow to stabilize. Reflective boundary condition was imposed at the walls and non-reflective boundary conditions were imposed at the inlet and outlet. Sound velocity of air was set to 340 m/s.

# CHAPTER 4: RESULTS AND DISCUSSION

## 4.1 Validation of Turbulence Models

Simulations were validated by comparing the mean axial velocities measured at 7 points on the centerline of the tube as shown in Figure. 4-1. Point P1 is 3 cm downstream from the constriction end and all measurement points are spaced 3 cm from the adjacent points.

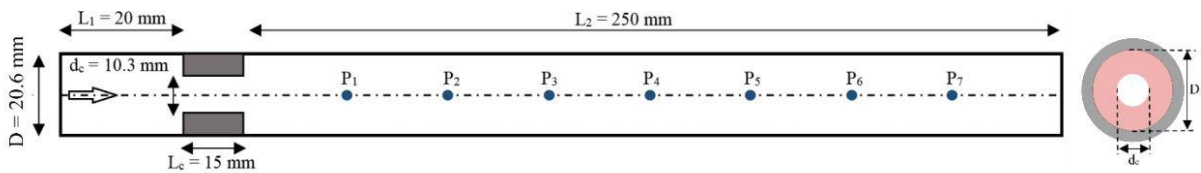

Figure 4-1: Velocity measurement locations in the tube

Figure. 4-2 shows the measured axial velocity signals at the 7 measurement points on the center of the pipe. Table 4-1 shows the mean and root mean square (RMS) value of the velocities at each point. At P1 and P2 high velocity values were observed where the high velocity jet has initiated due to the area reduction at the constriction. At P3 a significant drop in the velocity was observed compared to P2. Then the velocity values continued to decrease till P5 and slightly increased from P5 to P7. Based on RMS values of the velocity, velocity fluctuations increased from P1 to P3 and then decreased from P3 to P7. These results indicate the existence of a high velocity jet after the constriction (at P1, P2) which become unstable and gradually dissipates (at P3,P4,P5). Then the flow is stabilized (at P6, P7). Similar flow behavior was observed in previous experimental studies[8, 51].

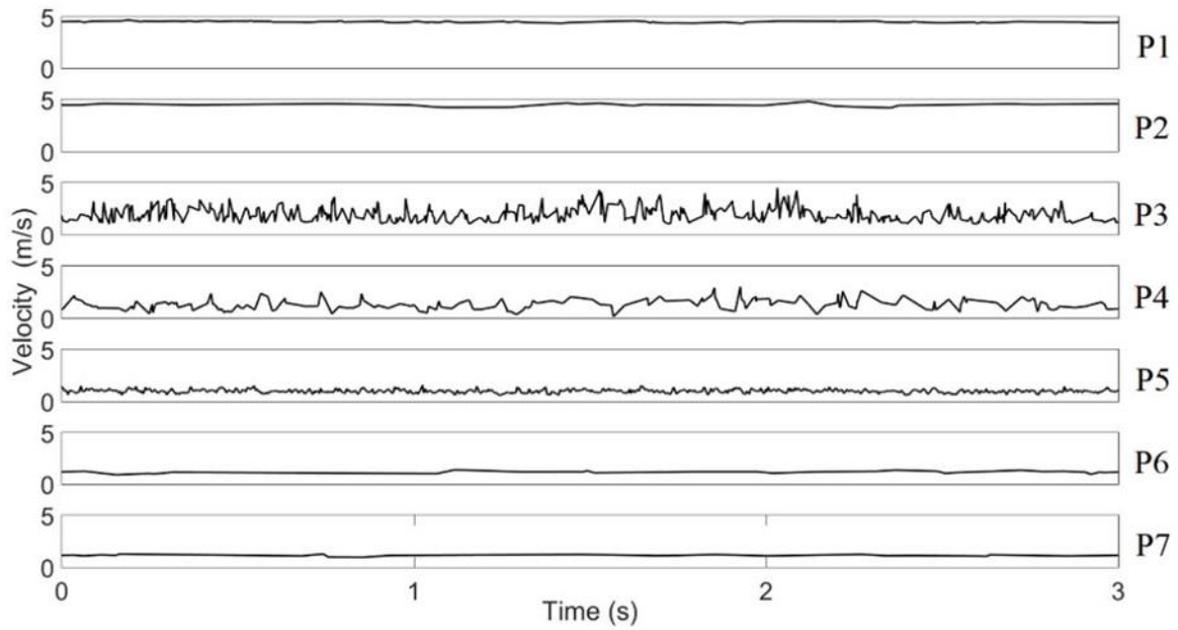

Figure 4-2: Measured axial mean velocities along the centerline of the tube

Table 4-1: Mean and RMS values of measured velocities

| Measurement point | Mean axial velocity (m/s) | RMS of axial velocity (m/s) |
|---|---|---|
| P1 | 4.66 | 0.047 |
| P2 | 4.63 | 0.115 |
| P3 | 2.09 | 0.708 |
| P4 | 1.17 | 0.503 |
| P5 | 1.11 | 0.205 |
| P6 | 1.14 | 0.126 |
| P7 | 1.16 | 0.063 |

Figure. 4-3 shows the comparison between the experimental and simulated results as well as the cross-sectional axial mean velocity distribution plots from the simulations. The results showed that LES model had the best agreement while DES SST k-ω model had the maximum error. RANS models had a better agreement compared to DES SST k-ω model, while RANS RST model delivered better results compared to RANS SST k-ω model. All turbulence models had a good agreement at points P1, P2 and P6, P7 which are in the regions where the high velocity jet initiates and where the flow stabilizes near the outlet. All turbulence models had their highest errors at points P3,P4,P5 where the high velocity jet is expected to become unstable and dissipate.

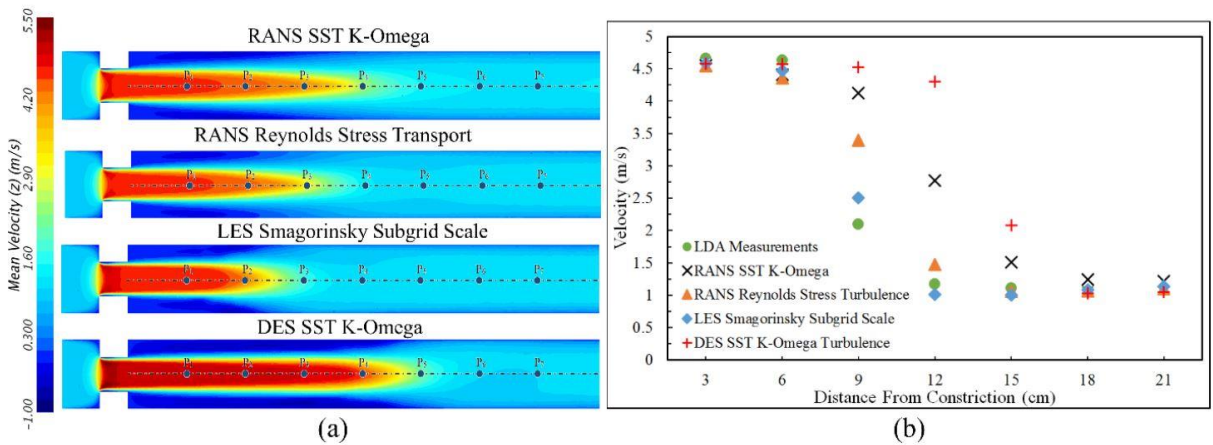

Figure 4-3: (a) Axial mean velocity plots at a cross-section of the pipe (b) Comparison of measured and simulated axial mean velocities along the centerline of the tube

As the objective of current study is to model the flow generated sound, accurate modelling of flow fluctuation is paramount to capture sound sources. Hence, RMS of velocity and instantaneous vorticity in the flow domains were compared between each turbulent model. Figure. 4-4 shows maps of RMS of velocity and vorticity at a cross section of the tube. Here, the vorticity values are displayed in the range of 0.1 s$^{-1}$ to 8000 s$^{-1}$ for clear comparison of vorticity fluctuations at the high fluctuation zone. The results clearly showed that RANS models perform poorly in capturing flow fluctuations, compared to LES and DES models. This is expected since RANS models solve for average flow quantities. Although, DES model could capture flow fluctuations, it predicted a delayed flow separation where high fluctuating zone moved toward the downstream. Not only LES model accurately captures the flow fluctuations, it also captured smaller eddies compared to DES as seen in vorticity results. As the DES model in the current study is a hybrid approach of SST k-ω and LES, inaccurate modelling of the flow near flow separation by SST k-ω is a probable reason for DES model over predicting the fluctuating zone towards downstream.

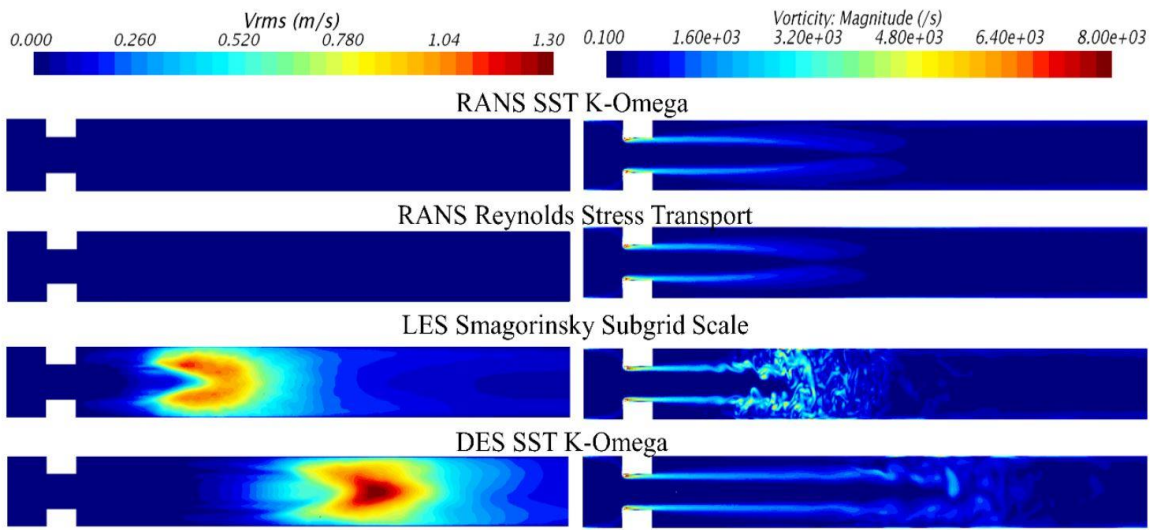

Figure 4-4: RMS of velocity fluctuations and vorticity at the flow domain cross-section

Based on CFD simulation results discussed above, LES was identified to deliver the most accurate results and was selected as the turbulence model to be used with CAA simulation.

Figure. 4-5 shows the distribution of axial velocity and streamlines inside the tube obtained from LES simulation. Here, the flow domain is classified in to different regions referring to previous studies [8, 51].

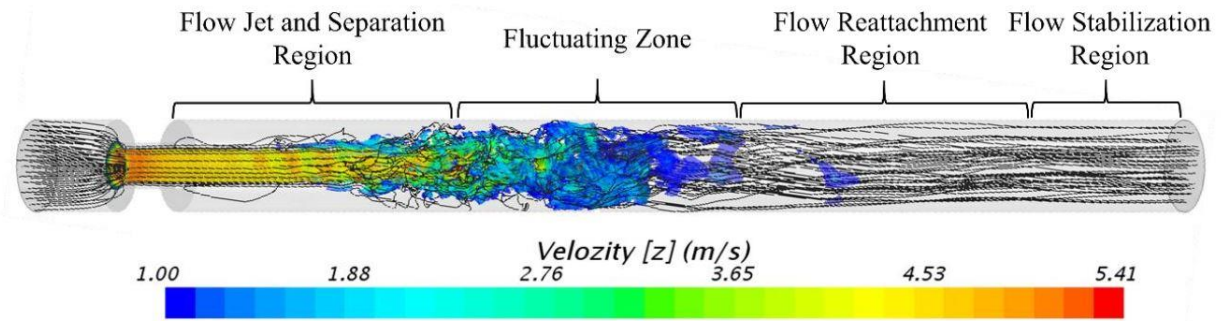

Figure 4-5: Distribution of axial velocity and streamlines inside the tube with region classification

### 4.2 Validation of Acoustic Simulation

4.2.1 Microphone Measurements

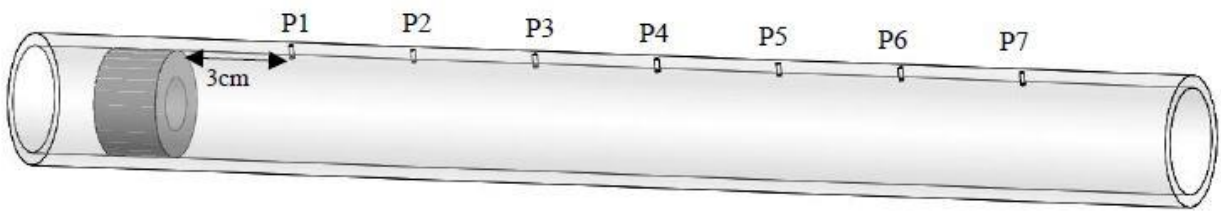

Figure 4-6: Pressure port locations for sound measurements

Sound was measured by connecting a microphone to the 7 pressure ports created on the wall tube. Pressure port p1 was placed 3 cm downstream the constriction and each port is placed 3cm spaced from adjacent ports. These port locations are selected to be consistent with the locations of the LDA measurements. Sound pressure level (SPL) was calculated using equation 74.

$$SPL = 20 \times \log\left(\frac{P'}{P_{ref}}\right) dB \qquad (74)$$

where, $P_{ref} = 2 \times 10^{-5}\ Pa$.

Figure. 4-6 shows measured sound pressure spectra at each pressure port including the background noise measured when there was no fluid flow inside the pipe. For all measurements, it can be seen that SPL rapidly decreases for frequencies less than 50 Hz. This is expected due to low sensitivity of the microphone for frequencies less than 50 Hz as observed in microphone calibration (see figure. 3-2). Hence, frequencies less than 50 Hz are not considered for validating the simulated results. While sound pressure spectra showed common frequency peaks (at 180 Hz,249 Hz, 420 Hz), they were not clearly visible for locations p3 and p4 which are located in the high fluctuation zone. Maximum sound pressure level was observed at p3. This was consistent with the LDA measurements where the highest fluctuations were at P3. In comparison with CFD results, it can be observed that high SPL levels are observed in the "fluctuating zone" and SPL levels reduce as the sound was measured away from the fluctuation zone.

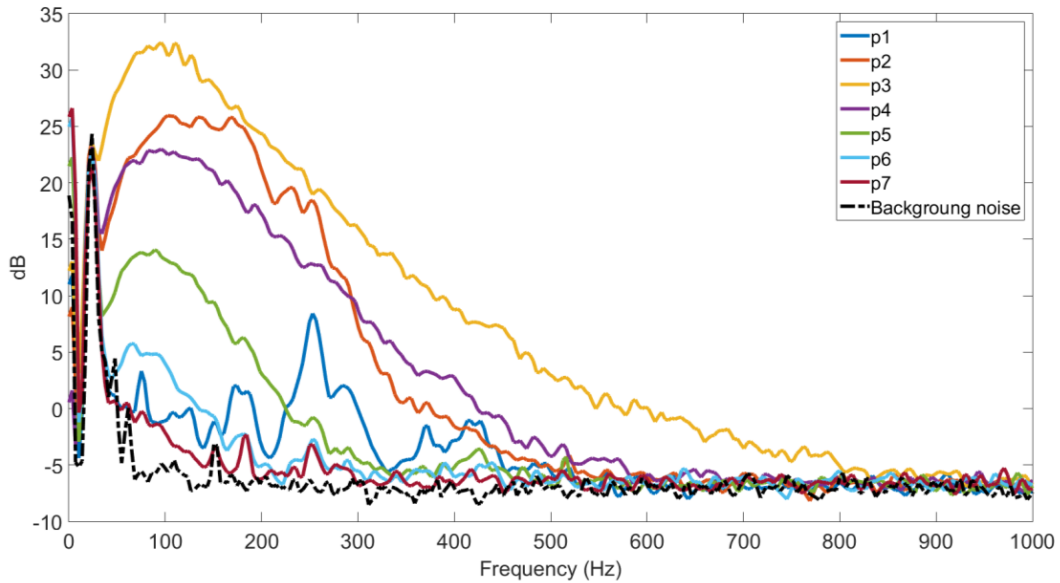

Figure 4-7: Measured sound pressure spectra

4.2.2 Comparison between Experimental and Numerical Sound Pressure Spectra

Figure 4-8 shows the comparison between measured and sound pressure spectra at each pressure port. Here, spectra for both total sound pressure and acoustic pressure calculated using APE method was compared (see section 2.2.2.3). Here the acoustic pressure represents the irrotational pressure field travelling at sound velocity while the total sound pressure consists of both hydrodynamic pressure and acoustic pressure. A microphone will measure total sound pressure in the near field. The comparison results showed a good agreement between the simulated total sound pressure and experimental sound pressure spectra. The spectra of measured sound pressure were shifted 30 dB up based on the microphone calibration results (see figure. 3-3).

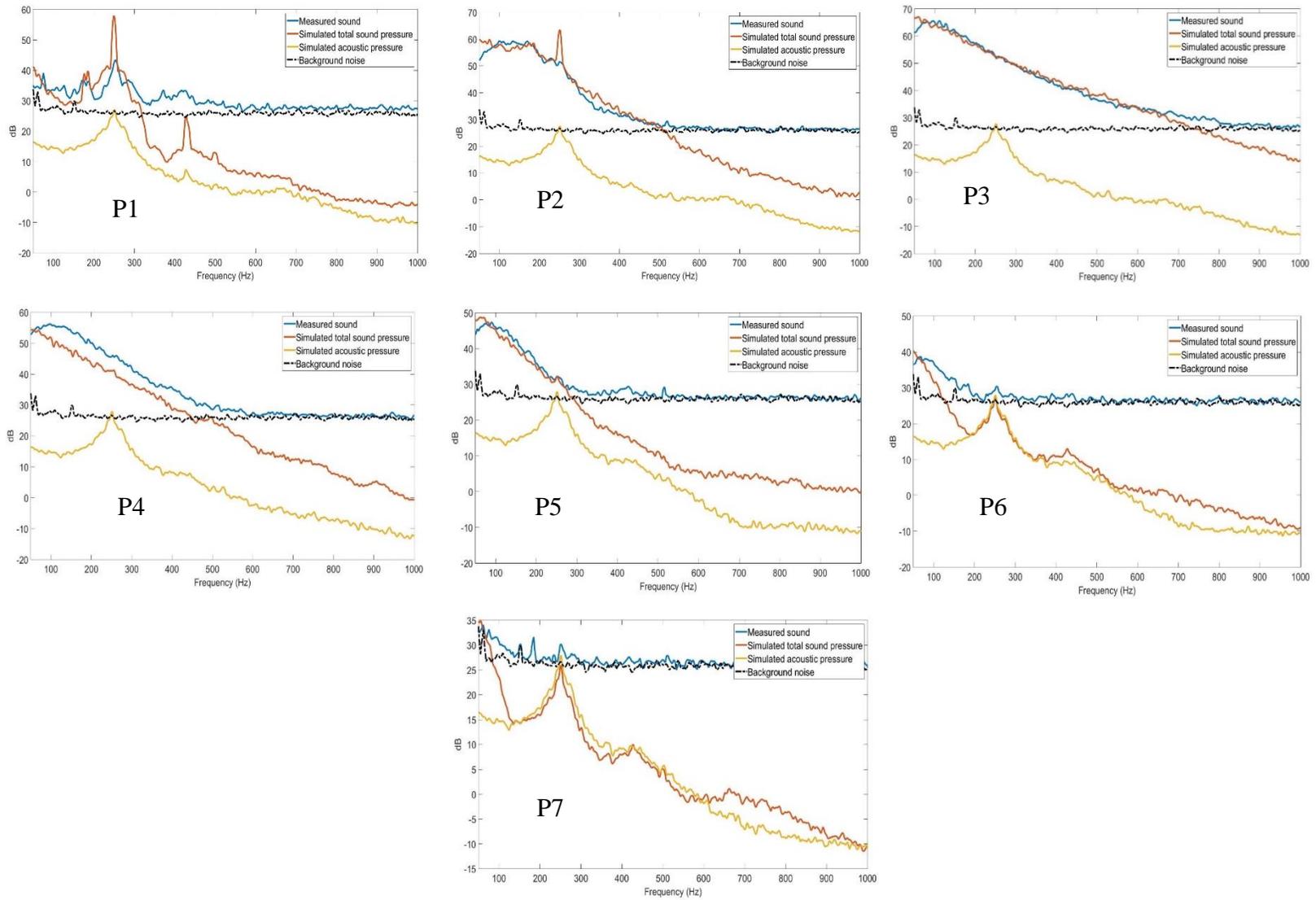

Figure 4-8: Comparison between measured and simulated sound pressure spectra

While clear distinct frequency peaks ( at 249 Hz and 420 Hz) were visible in the sound pressure spectra at the measurement points p1, p5,p6,p7 ,spectra at p2,p3,p4 had a broadband distribution without clear peaks. This may indicate the generation of broadband frequencies in the fluctuating zone (p2, p3, p4 are located in the fluctuating zone). These frequencies damp as they move away from the fluctuating zone, giving rise to clear frequency peaks observed in the sound pressure spectra measured away from the fluctuating zone. A significant difference between the simulated total sound pressure and acoustic pressure was observed in the fluctuating zone and this difference decreased for measurements taken away from the fluctuating zone. This indicates the high contribution of the hydrodynamic pressure (pseudo sound) to SPL in the fluctuating zone and it's decay as the sound travels away from the source region.

## 4.3 Acoustic Sources and Propagation

The source terms which excites the acoustic field in APE method is $\frac{1}{c^2}\frac{\partial^2(P')}{\partial t^2}$ (see equation 58). In the work by Ewert & Schroter (ref) This source term is described as a vortex sound source which generates sounds due to vortical fluctuations of the flow. However, it should be noted that this source term is responsible for perturbation of acoustic pressure $p^a$ which is described as the perturbation with hydrodynamic perturbations (pseudo sound) excluded. Here, $p^a$ can be described as an irrotational field (with no vortical structures) which can be low in the source region with vortical fluctuations for low Mach number flow [35]. The hydrodynamic perturbations are highest in the source region and will disappear in far field. In the current study, as the analysis is contained to the near field, the total pressure perturbation which consists both hydrodynamic and acoustic perturbation ($p^a + P'$) is considered as sound.

### 4.3.1 APE Sources

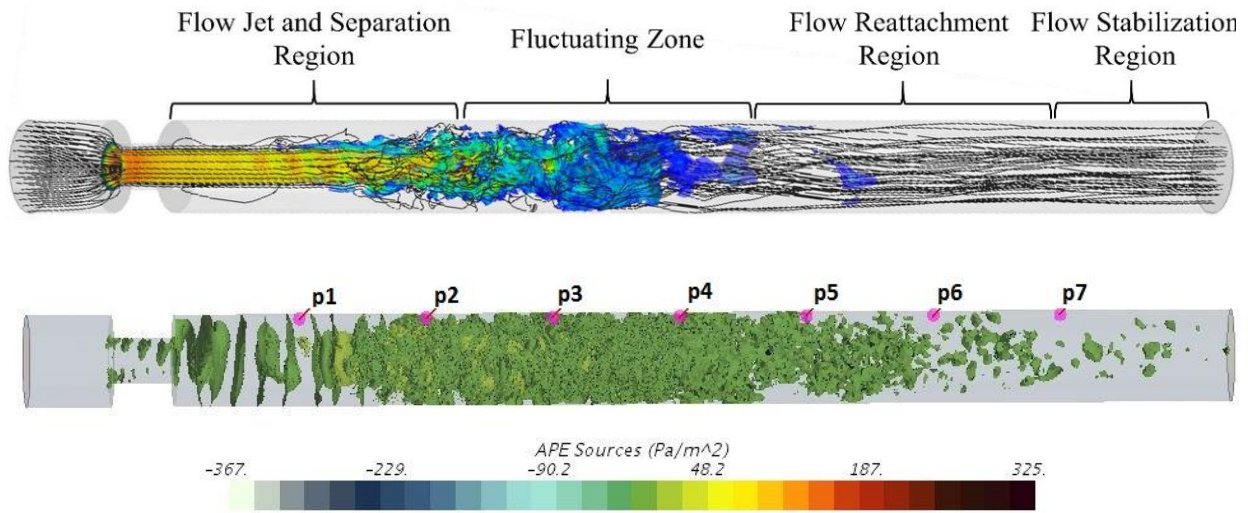

Figure 4-9 : (top) Flow regions (bottom) Volume representation of APE sources at t=15 s using 60 Iso-surfaces

Figure 4-9 shows a volume representation of APE acoustic sources. Organized, ring-like APE source regions can be identified close to the constriction. These organized regions become disoriented as they move further downstream and disappears close to the outlet. For, further investigation of the frequencies of acoustic sources, surface FFTs of the APE sources were plotted for different frequency values. Figure 4-10 shows surface FFTs of APE sources at several frequencies of 180 Hz, 249 Hz and 420 Hz corresponding to the peaks observed in sound pressure spectra. The surface FFT results showed that APE sources fluctuating at 249 Hz were concentrated in flow jet and separation region, between p1 and p2 (Figure 4-9), while APE sources for 180 Hz were visible at the end of flow jet and separation region with low amplitudes compared to 249 Hz. APE sources for 420 Hz distributed in the fluctuating zone. High

amplitudes in the surface FFTs were observed in same regions for velocity fluctuations and vorticity ( see Appendix A).

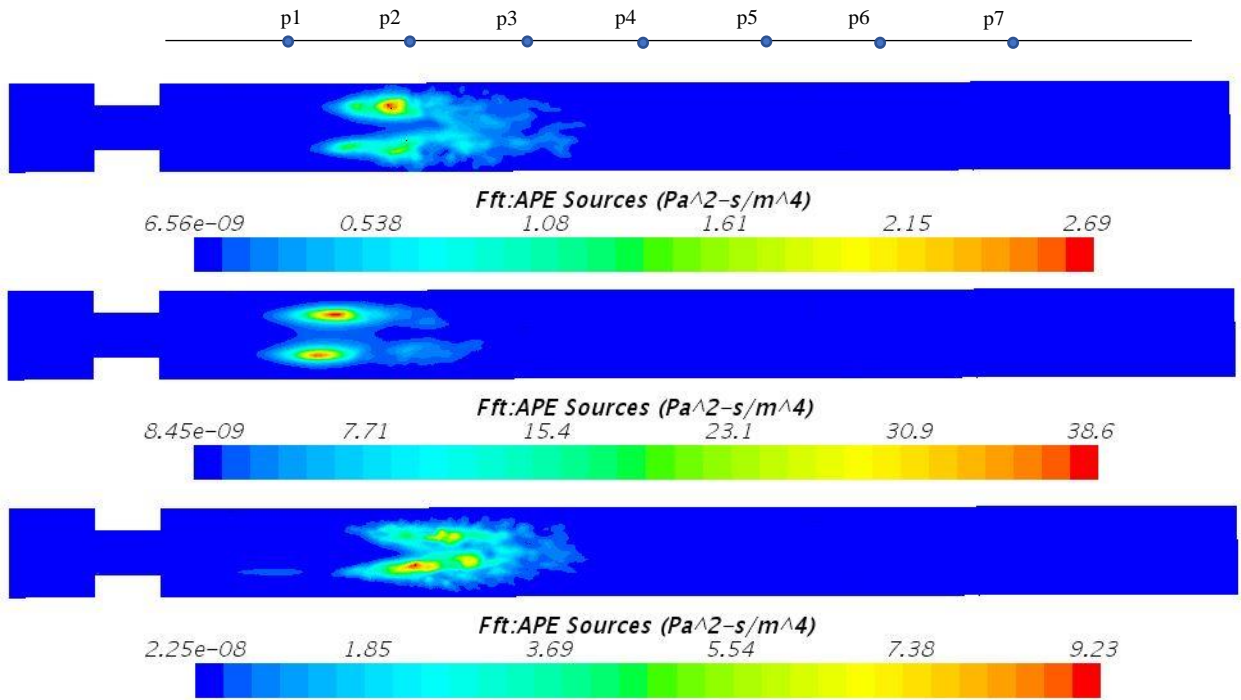

Figure 4-10: Surface FFTs of APE sources for frequencies 180Hz,249 Hz,420 Hz from top to bottom

4.3.2 POD Analysis

POD modes were calculated for flow variables, total sound pressure $p' = P' + p^a$. A snapshot matrix was saved for 1.07 second of data using a time step of 0.001 seconds. Data were saved for a volumetric region near the constriction (region containing p1,p2,p3). The snapshot matrix consisted of 1084167 rows and 1070 columns. Figure 4-14 shows highest energy mode (1st POD mode) of $p'$. By analyzing the variation of time coefficients of this POD mode, it was observed that mode frequency is 249 Hz which is a peak frequency of the sound pressure spectra measured on the wall (see Figure 4-8).

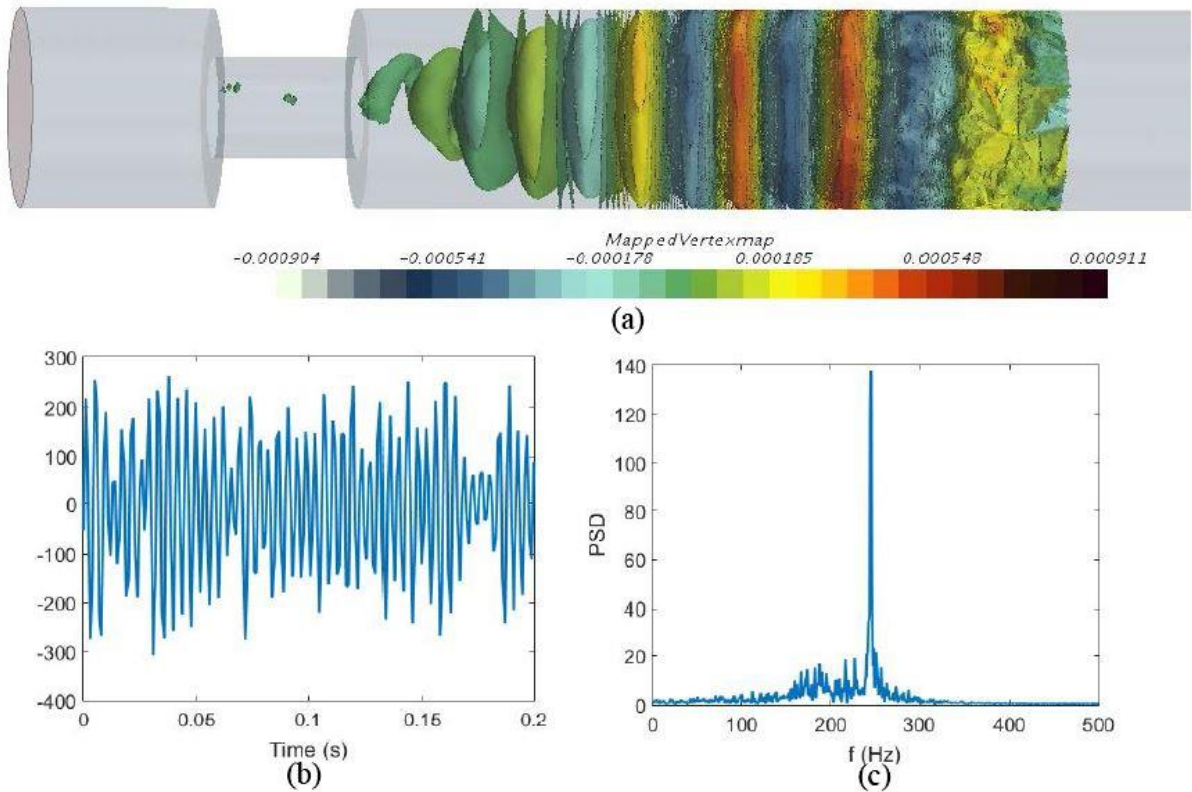

Figure 4-11: (a) Volume representation of POD mode 1 of total sound pressure (b) Variation of time coefficients of POD mode 1 (c) Spectrum of time coefficients f POD mode 1

According to the POD mode 1, the highest sound pressure at 249 Hz was observed in between measurement points p2 and p3, which is in the beginning of the fluctuating flow region where ring like coherent regions with maximum and minimum values adjacent to each other.

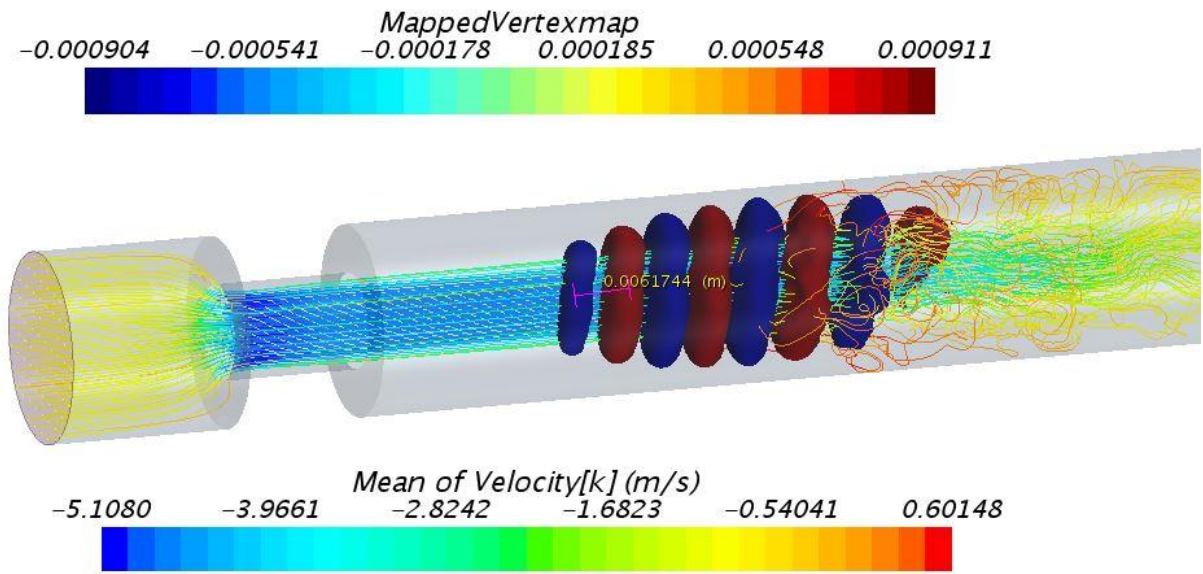

Figure 4-12 : Isometric view of the total sound pressure POD mode 1(using Iso-surfaces of minimum and maximum values) with velocity streamlines

Figure 4-12 shows the existence of ring-like mode structures in the region where the velocity jet starts to separate. Surface FFTs of Acoustic sources, velocity and vorticity at 249 Hz showed high amplitudes in the same region where the jet separation occurs. This is an indication that the frequency peak at 249 Hz is generated due to the flow fluctuations in the shear layer of the flow

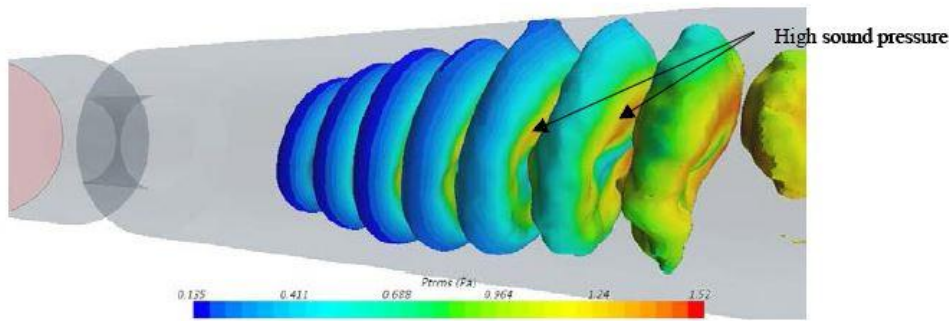

Figure 4-13: RMS of total sound pressure distribution plotted on the Iso-surfaces of POD mode 1

separation region of the high velocity jet , downstream the constriction. As the flow reaches further downstream and generate high fluctuations, the ring like mode structures are disorganized. The distance between these ring like structures were estimated as 6 mm (distance between two adjacent rings with similar color in Figure 4-12) . Considering the frequency they oscillate; the characteristic velocity of the POD structures was calculated using equation 75.

$$f = V/L \qquad (75)$$

where, $f$ is the mode frequency 249 Hz and $L$ is the characteristic length (similar to wave length) found as 6 mm. Then, characteristic velocity is found to be 1.49 ms$^{-1}$. This velocity value is in the same range of the mean convective axial velocity at the flow separation region (see Figure.4-14).

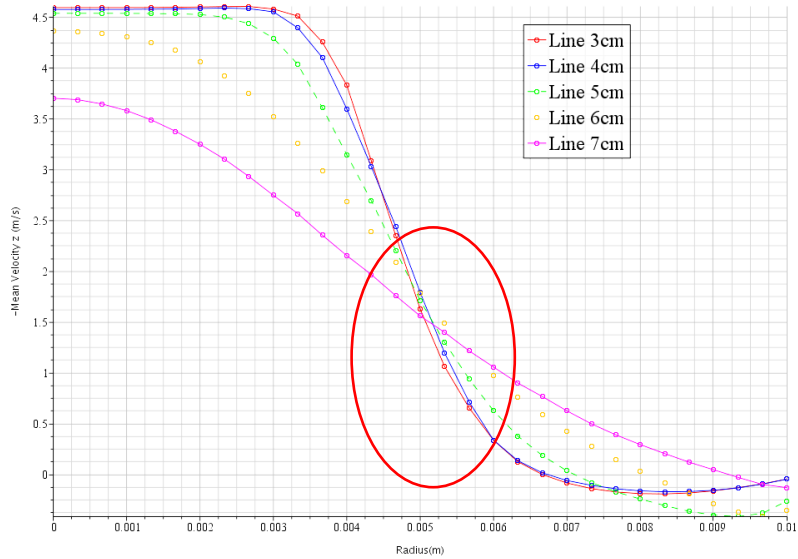

Figure 4-14 : Axial velocity profiles plotted at 3 cm, 4 cm, 5 cm, 6 cm and 7 cm downstream the stenosis

The similar velocity values in the jet separation region (shear layer) indicate that acoustic sources of 249 Hz are possibly due to flow fluctuations (or eddies) in the jet shear layer which travels at convective flow velocity.

Figure 4-13 shows the distribution of total sound pressure on the POD mode 1 Iso-surfaces. This indicates the radial propagation of total sound pressure towards the walls. The positive and negative values of the adjacent ring like structures observed in POD mode 1(see Fig 4-11 and 4-12), indicates the axial propagation of the total sound pressure towards upstream direction. POD modes were not plotted further downstream region due to computational limitations. Spectra of higher POD modes were broadband and weren't considered for analyzing the acoustic sources and propagation For an example see Appendix A for plots of POD Mode 6.

### 4.3.3 Surface FFT of Sound Pressure

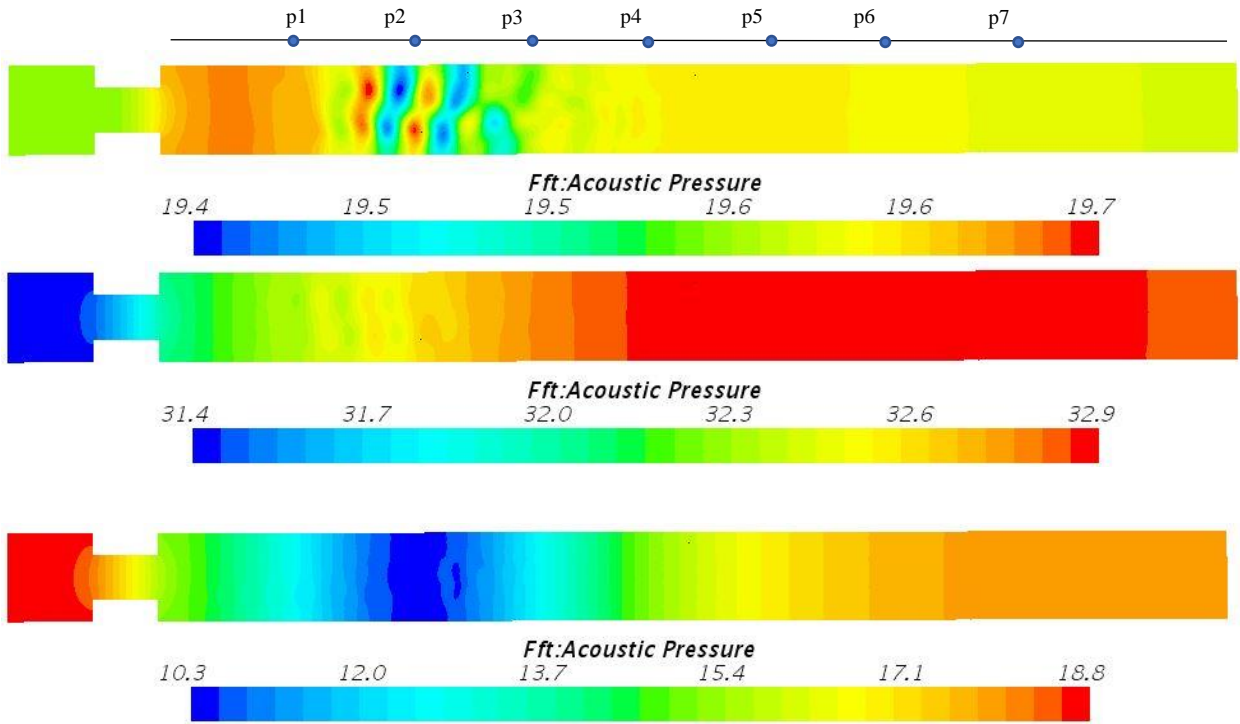

Figure 4-15 : Surface FFTs of acoustic pressure $p^a$ for frequencies 180Hz, 249 Hz, 420 Hz from top to bottom

Figure 4-15 and Figure 4-16 show surface FFTs acoustic pressure ($p^a$) and total sound pressure ($p^a + P'$) for different frequency ranges to study the sound propagation inside the tube at different frequencies. The amplitudes of the sound pressure levels are presented in dB.

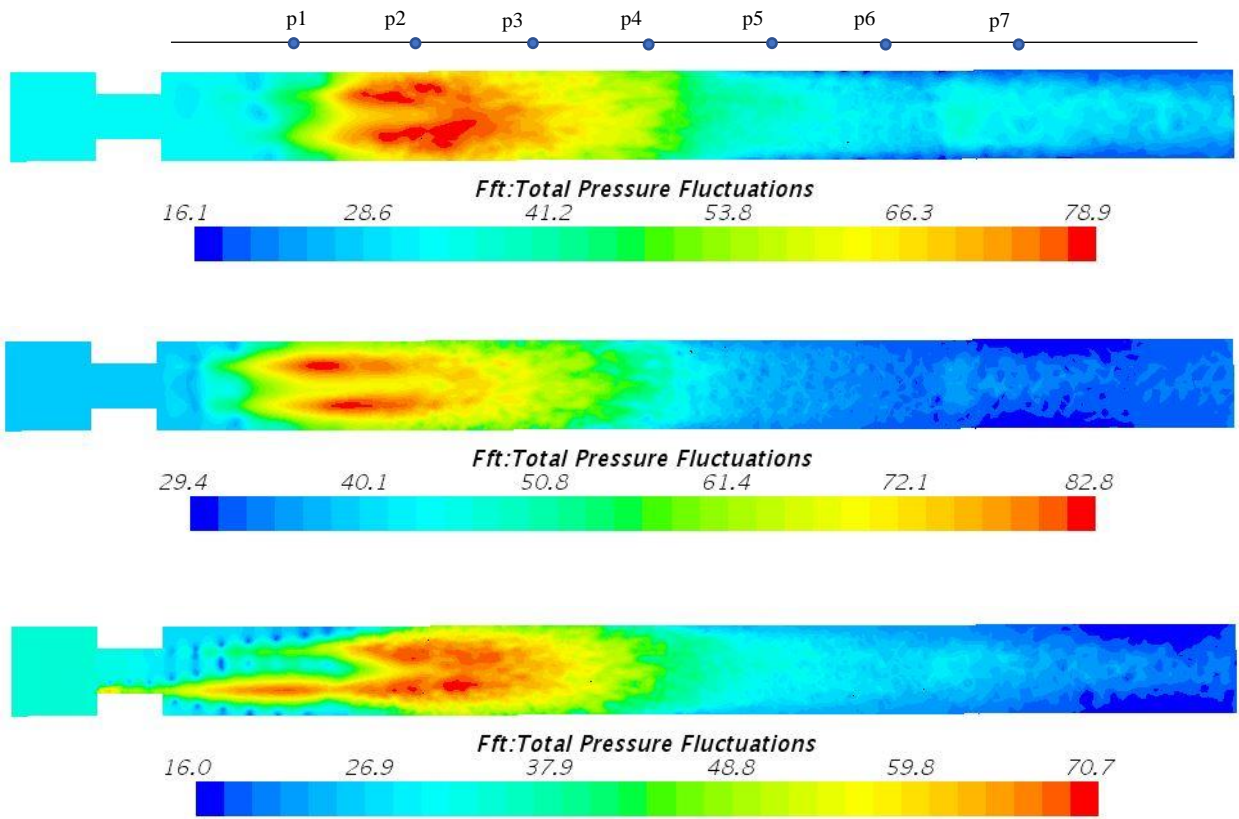

Figure 4-16 Surface FFTs of total sound pressure for frequencies 180Hz, 249 Hz, 420 Hz from top to bottom

Surface FFTs for irrotational acoustic pressure showed an axial wave propagation pattern and the amplitudes values didn't vary significantly inside the domain. In contrast, surface FFTs of total sound pressure showed very high amplitudes in the source regions which are significantly damped as they move away from the source region.

# CHAPTER 5: SIMULATING FLOW AND ACOUSTICS IN HUMAN AIRWAY MODEL

## 5.1 Airway Geometry

A realistic human lung airway geometry was extracted using computed tomography (CT) images using ITK-Snap software [52]. The geometry was exported as a surface mesh file (in. STL format) which was later edited in StarCCM+ to remove bad quality cells. The airway geometry included the upper airways including mouth and nasal passages. Lower part of the airway consisted of trachea and intrathoracic airways up to few generations.

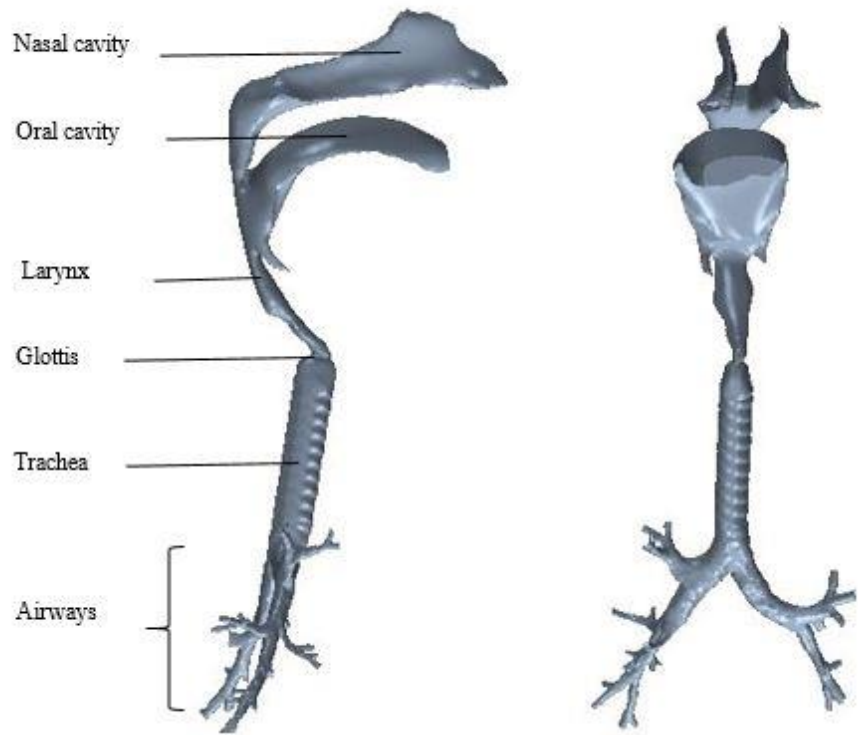

Figure 5-1: Realistic human airway geometry

## 5.2 Simulation Setup

### 5.2.1 Mesh

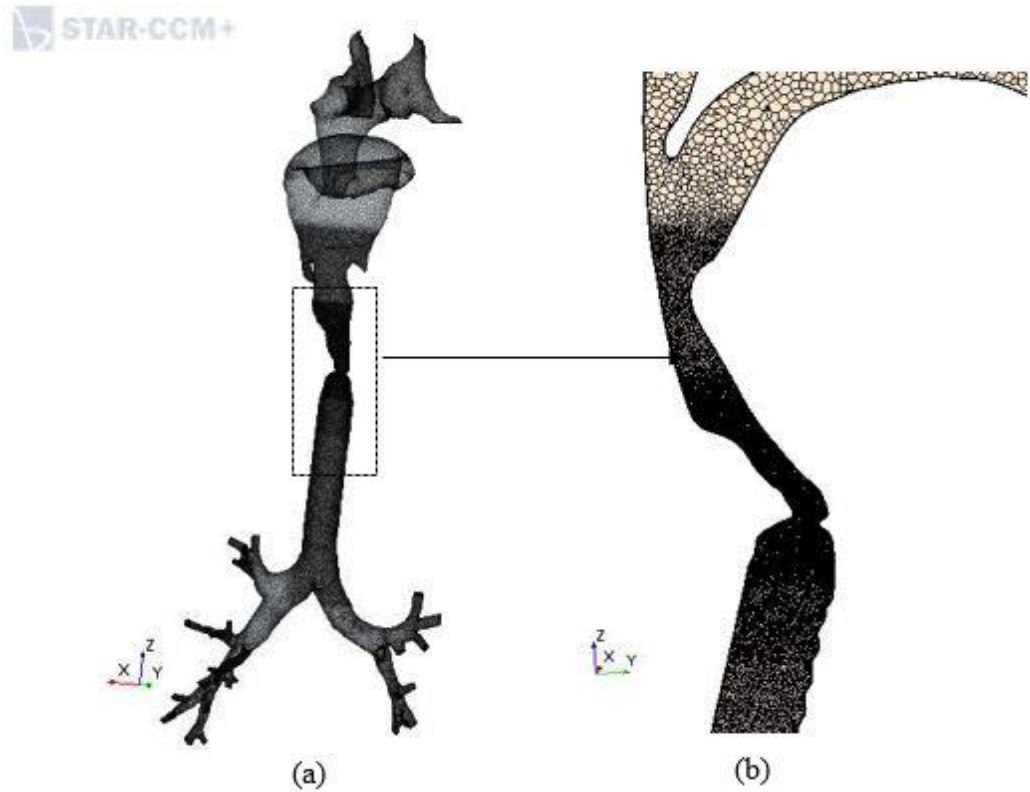

Figure 5-2: (a) CFD mesh of the human airway (b) Cross sectional view of the refined mesh region

The final mesh contained ~2 Million polyhedral type cells. The mesh was refined near the glottis where high fluctuations are expected as shown in Figure 5-2 (b). The average mesh size in the refined region was 0.25 mm. A prism layer mesh of 4 layers, with a total thickness of 0.3 mm and a stretching factor of 1.4 was used at the boundaries to maintain the Y+ value in the order of 1.

### 5.2.2 CFD Simulation

LES turbulence model with Smagorinsky SGS model was used to simulate the flow field. Time step was set to 1E-4 s. The density and the dynamic viscosity of the air was set to 1.184 kgm$^{-3}$ and 1.855 E-5 Pa.s, respectively. Mass flow rate boundary conditions was set at the mouth inlet. Here, the inlet flow rate corresponds to the peak inspiratory flow rate Zero static pressure boundary condition was imposed at the airway outlets and non-slip boundary condition was set at walls. Nose was neither considered as an inlet nor is outlet, assuming airflow through the nose blocked using nose clips as done in experimental procedures.

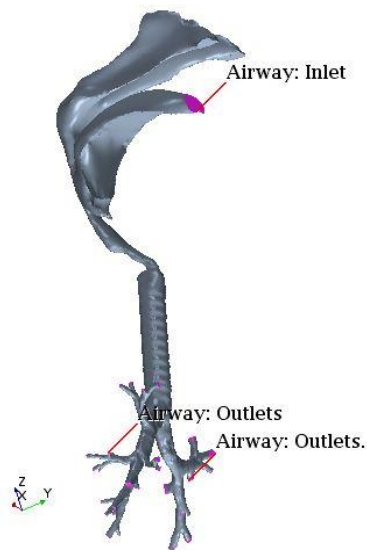

Figure 5-3: Airway CFD boundary conditions

### 5.2.3 CAA Simulation

CAA simulation was performed using APE based hybrid method. Time step was set to 0.0001 s. Non-reflective boundary conditions were set at the inlet and outlets. Reflective boundary condition was imposed at walls including the nasal cavity. CAA solver was started after 3 seconds, allowing the CFD solution to stabilize.

## 5.3 Results and Discussion

### 5.3.1 Flow Results

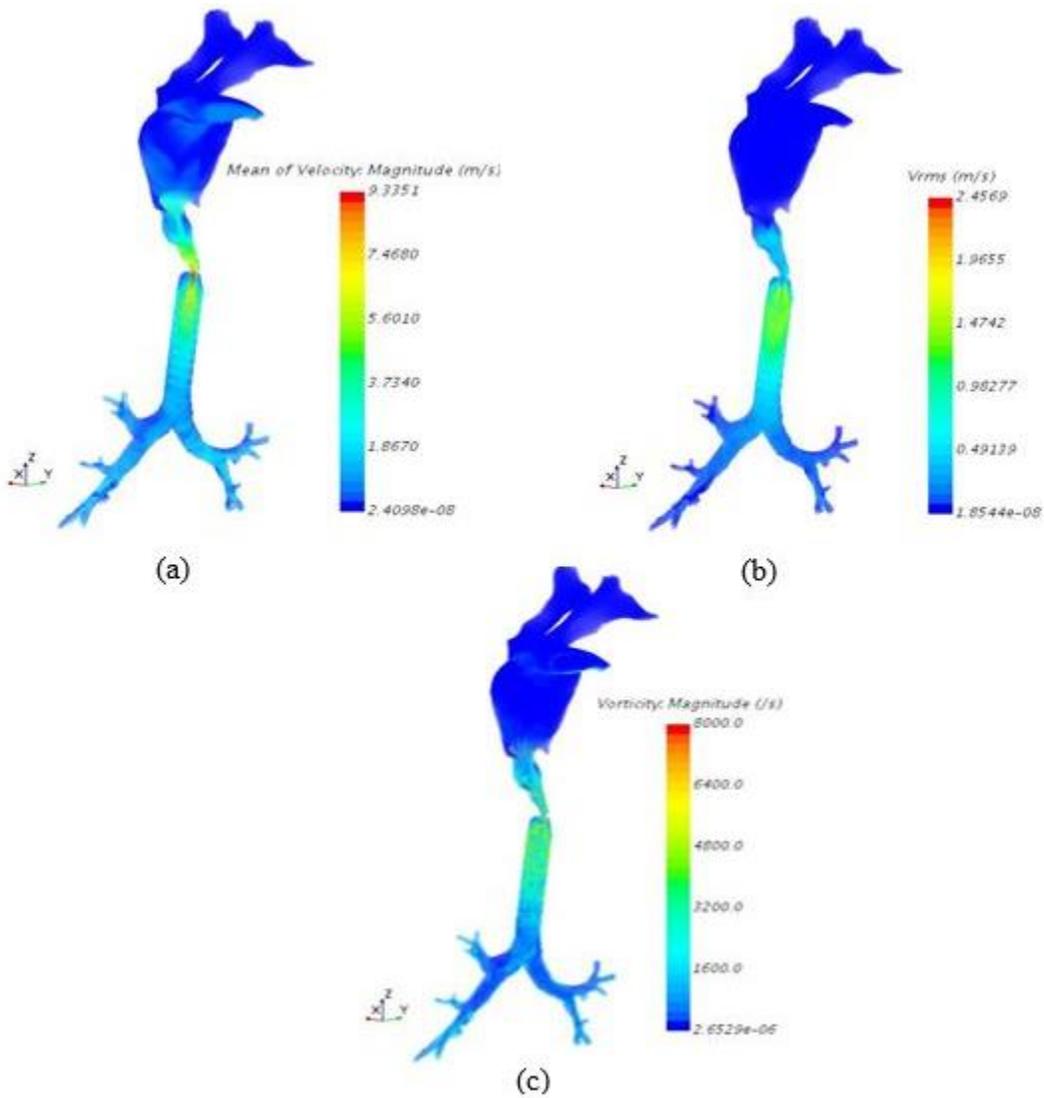

Figure 5-4: Volumetric representation of flow results for an inlet flow of 0.32 l/s (a) Mean velocity distribution (b) RMS of velocity (c) Instantaneous vorticity

Maximum velocity was observed at the glottis as expected (see Figure 5-4 (a)). As shown in Figure 5-4 (b), highest flow fluctuations were observed in the trachea, downstream the glottis. Considerable flow fluctuations were observed upstream the glottis close to larynx. Comparable results were observed for vorticity distribution. The velocity results also showed the impingement of velocity jet formed at the glottis on the anterior wall of the trachea, due to the angle between trachea and upper airways (See Figure 5-5)

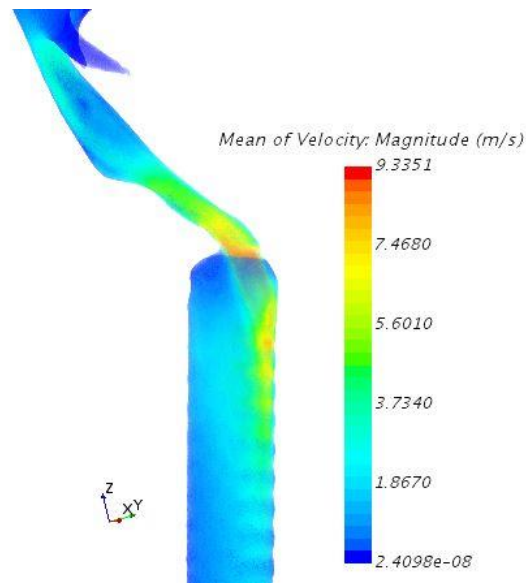

Figure 5-5: Impingement of the velocity jet on anterior tracheal wall

. The mean mass flow ratios in to left and right lung airways was calculated as 46% and 54% of the inlet flow, which are in agreement with previous studies [53, 54]. The pressure drop between the inlet (mouth) and outlets (airway outlets) was calculated as 49 Pa.

5.3.2   Acoustic Results

Figure 5-6 shows the results for APE sources. Highest amplitudes were observed downstream the glottis, in high fluctuation region where the velocity jet impinges on the tracheal wall.

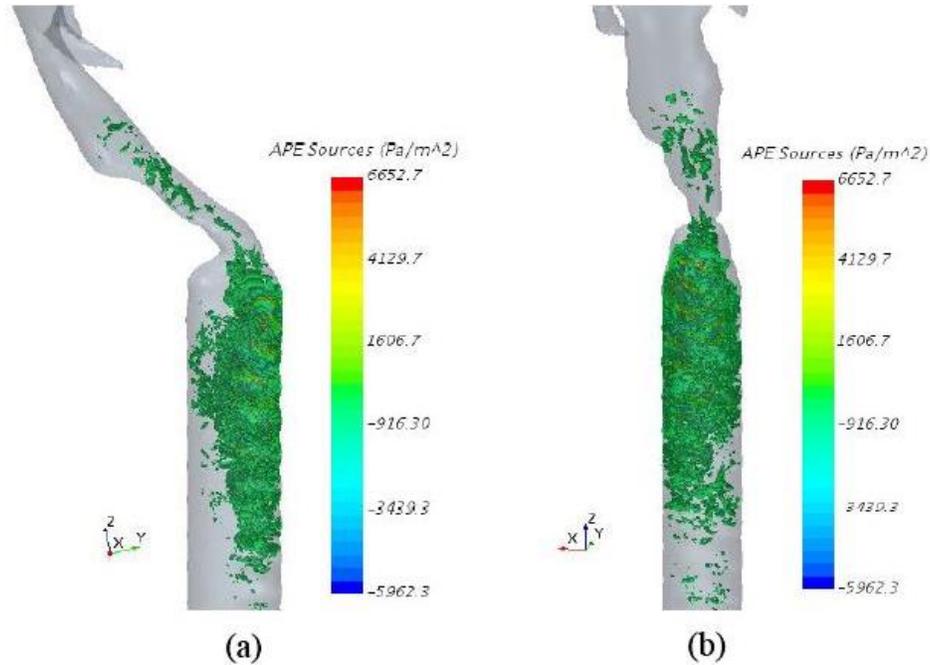

Figure 5-6: APE sources in human airway using 60 Iso-surfaces for inlet flow 0.32 l/s(a) Side view (b) Front view

Figure 5-7 shows the distribution results for RMS of total sound pressure $(p' = P' + p^a)$ and irrotational acoustic pressure $p^a$. The highest total sound pressure was observed downstream the glottis near the anterior tracheal wall. In contrast, highest acoustic pressure $(p^a)$ was observed at the edge of nasal passages. This may be due to the reflections of irrotational acoustic waves on the narrow nasal passage walls with relatively sharp corners. The effect of high acoustic pressure in the nasal passage is also visible in total sound pressure results.

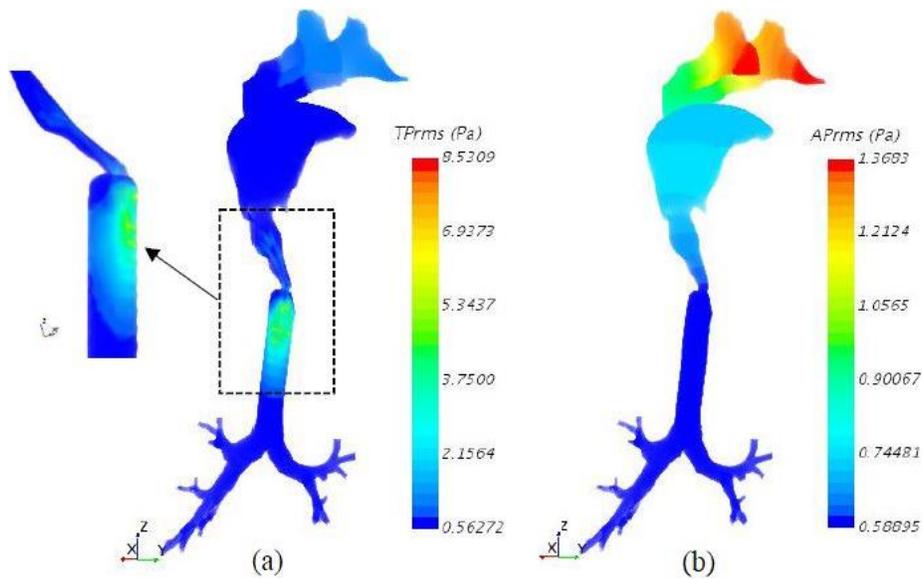

Figure 5-7: CAA sound pressure results for human lung airway for inlet flow 0.32 l/s (a) Total sound pressure (b) Irrotational acoustic pressureAcoustic pressure spectra and total sound pressure spectra were plotted at different probe points in the airway geometry.

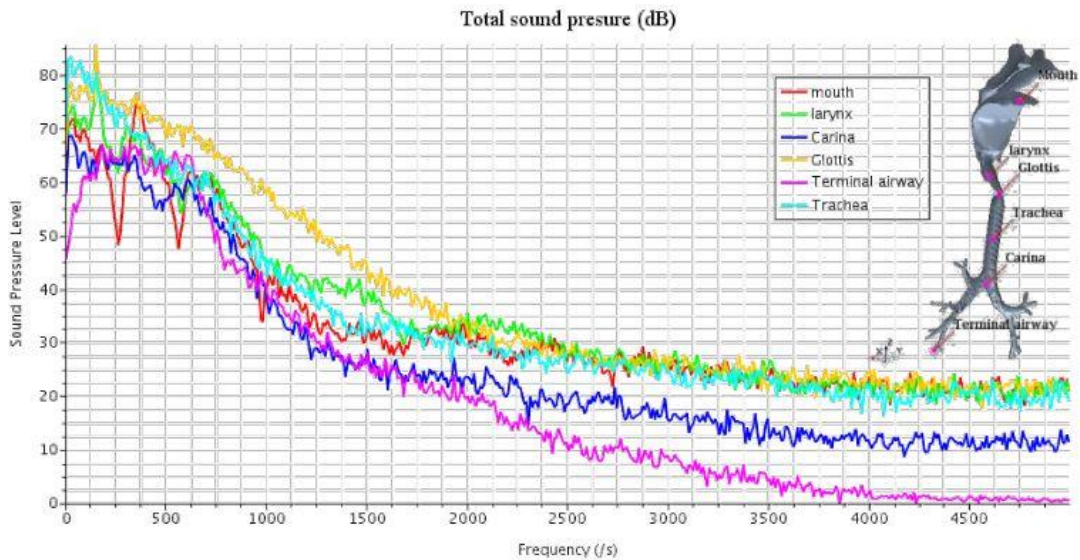

Figure 5-8: Total sound pressure spectra measured at different points in airway with inlet flow 0.32 l/s

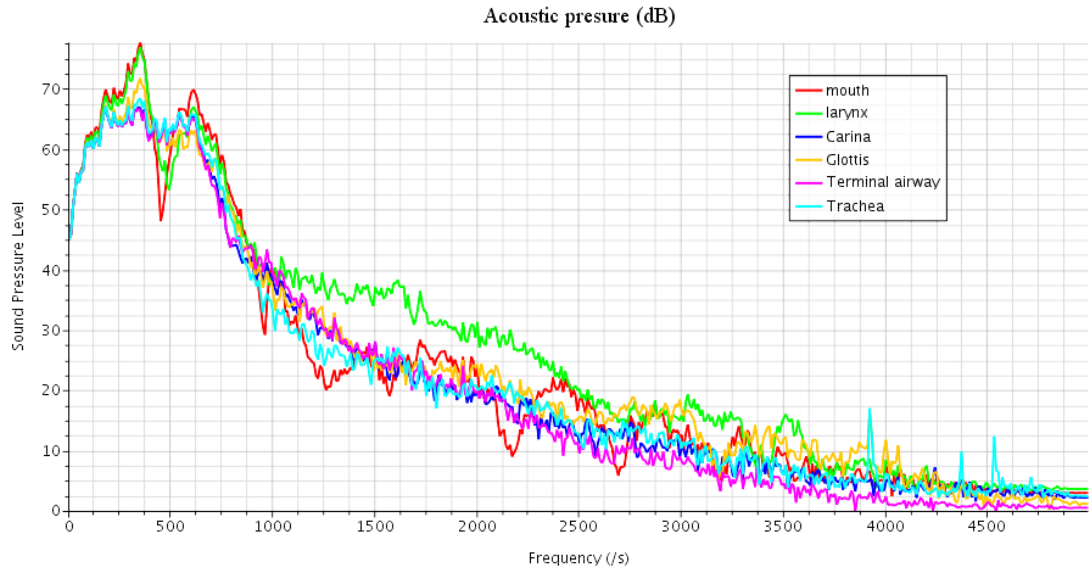

Figure 5-9 Acoustic pressure spectra measured at different points in airway for inlet flow 0.32 l/s

In contrast to total sound pressure spectra, two clear frequency peaks were observed at 350 Hz and 600 Hz in the acoustic pressure spectra for all measurement locations. While relatively high total sound pressure amplitude was observed at the glottis, highest acoustic pressure amplitudes were observed at mouth (between 0-900Hz) and larynx (between 1000-2500Hz).

### 5.3.3 Variation of Acoustic Power with Inspiratory Flow Rate

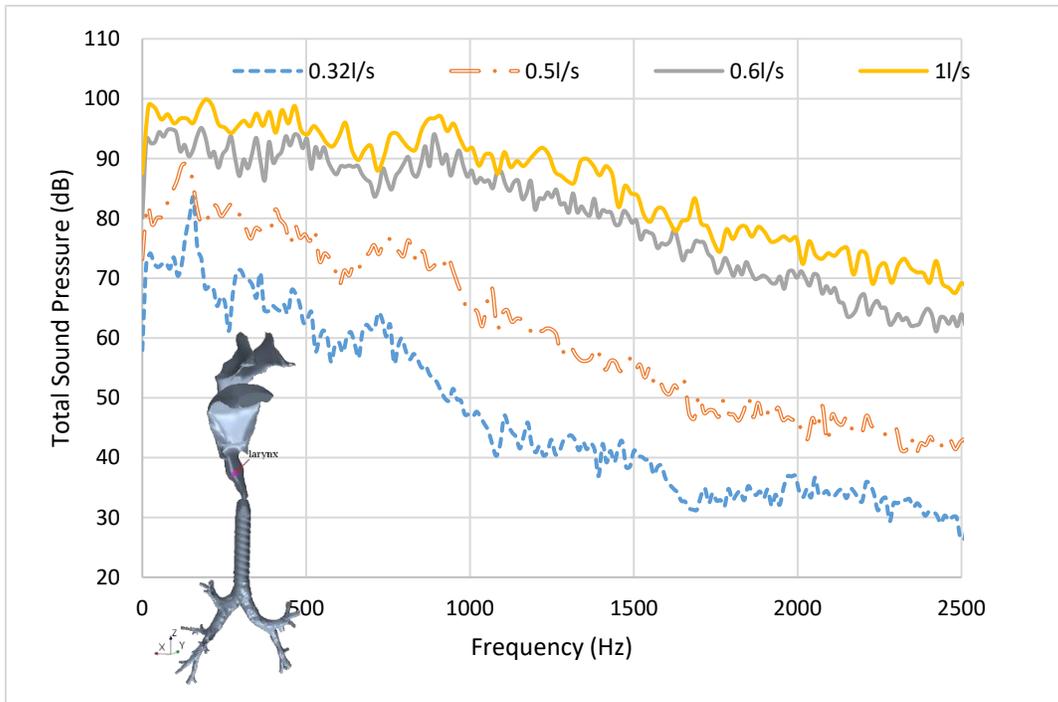

Figure 5-10: Total sound pressure spectra on larynx wall measured at different flow rates

Total sound pressure was recorded at the mouth for different flowrates of 0.32 l/s, 0.5 l/s, 0.6 l/s and 1l/s. Sound amplitudes increased with the increase in flowrate. Frequency peaks also slightly shifted with the increase of flowrate. The total sound power was calculated for each flow rate by calculating the area under the spectrum. The results showed that the increment rate of sound power decreases with the increase of flow rate. Comparable results were observed in a previous experimental study [55] as shown in Figure 5-12.

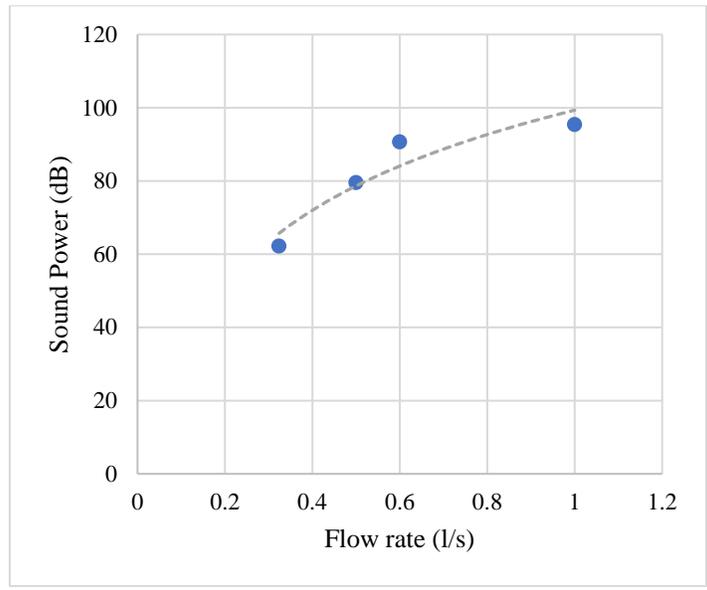

Figure 5-11: Sound power vs. flow rate for 300 Hz-600 Hz

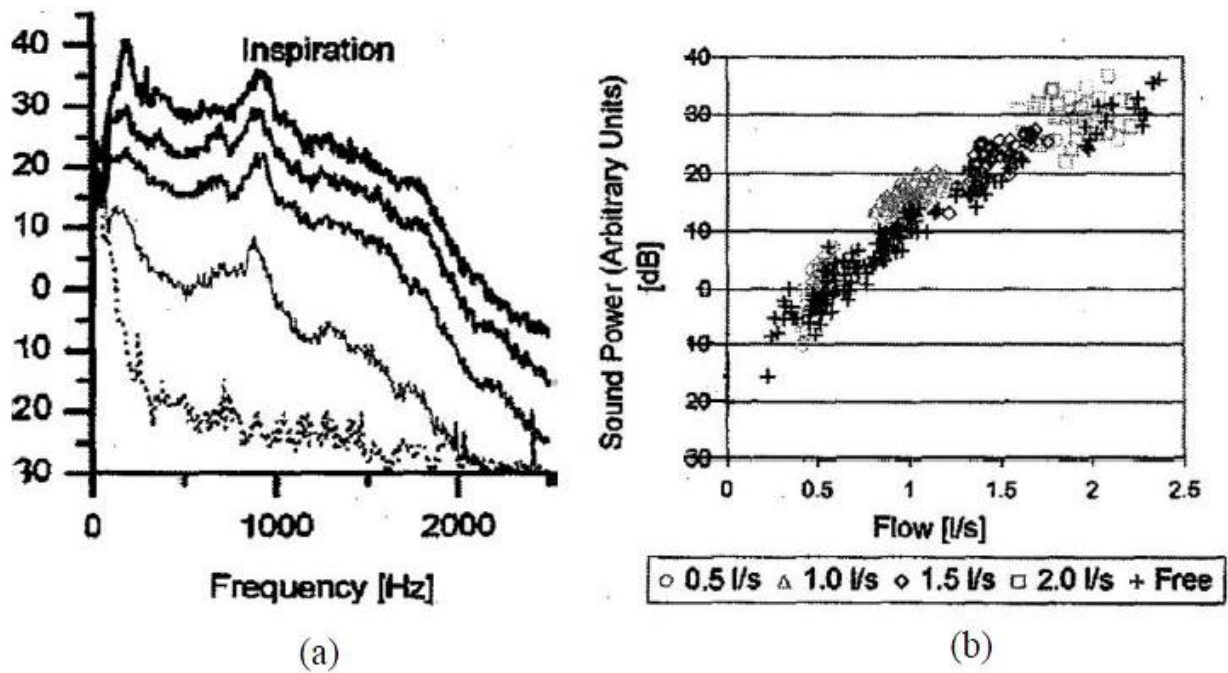

Figure 5-12: Experimental measurements of inspiratory breath sounds [54] (a) spectra for different flow rates (b) variation of sound power between 300-600Hz with flow rate

Another, study was carried out to examine the validity of a theory derived by Meyer-Eppler (1953) [49] related to the sound generation in vocal tract. Meyer-Eppler used constricted plastic tubes to model vocal tract and measured sound pressure at the mouth inlet at different flow rates. Based on the measurements following equation was derived.

$$P_{rms} = K(Re^2 - Re_{crit}^2) \tag{76}$$

where, $P_{rms}$ is the root mean square of the sound pressure measured at the inlet and $Re$ is the Reynolds number (measured at the glottis constriction). $Re_{crit}$ was defined as the critical Reynolds number where sound is initially detected when increasing the flow rate. $K$ is a constant possibly depends on fluid properties and flow domain geometry. Equation 76 can be further simplified to Equation 77, considering the constant properties of the fluid and the geometry. Here, $K'$ is a constant which includes the fluid properties density and viscosity. $V$ is the mean velocity at the glottis and $V_{crit}$ is the velocity corresponds to $Re_{crit}$ at glottis.

$$P_{rms} = K'(V^2 - V_{crit}^2) \tag{77}$$

Table 5-1: Mean velocity at glottis and RMS sound pressure at mouth inlet at different inspiratory flow rates

| Flow rate (l/s) | V (m/s) | V² | Prms (Pa) |
|---|---|---|---|
| 0.32 | 8.315033762 | 69.13978646 | 0.374157 |
| 0.5 | 12.99224025 | 168.7983068 | 1.037596 |
| 0.6 | 15.5906883 | 243.0695618 | 1.430642 |
| 1 | 25.9844805 | 675.1932271 | 4.40763 |

Table 5-1, shows the mean velocity at the glottis and $P_{rms}$ at different inspiratory flow rates. Using these results, $P_{rms}$ was plotted against $V^2$ to examine the validity of Equation 77 for current application.

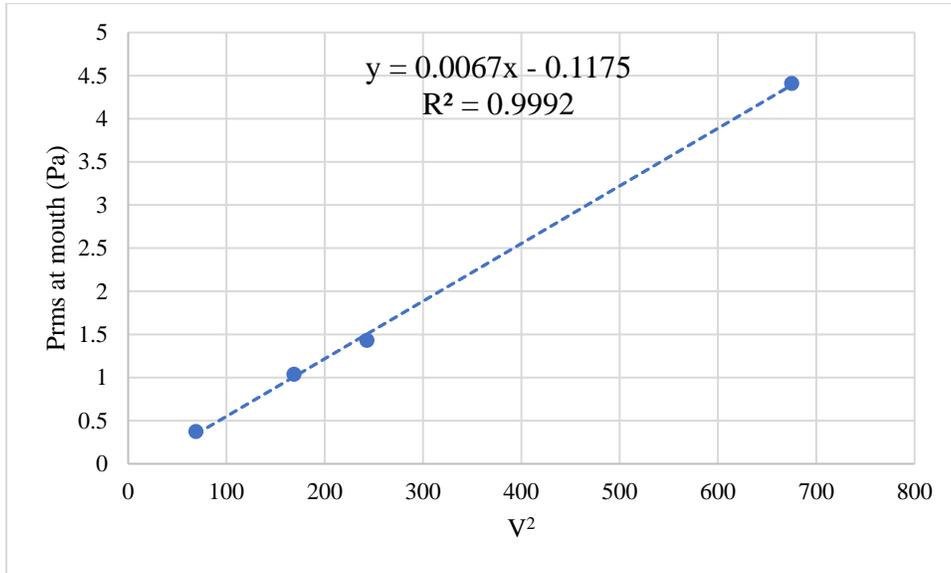

Figure 5-13: Variation of RMS of sound pressure at mouth inlet with the square of mean velocity at glottis

As shown in Figure. 5-13, results showed a good agreement with Equation 77. Based on the theory suggested by Meyer-Eppler (1953), $V_{crit}$ can be calculated as 4.187 m/s which corresponds to an inspiratory flow rate of 0.161 l/s .

# CHAPTER 6: CONCLUSION AND FUTURE WORK

The flow field and acoustics in a constricted circular duct at low Mach number is numerically studied and validated. Different turbulence models were used to simulate the flow field and results were validated by comparison with LDA velocity measurements. Turbulence models included RANS SST k-ω, RANS RST, DES SST k-ω and LES with Smagorinsky SGS model. Flow results showed that LES had the best agreement with LDA measurements and could capture velocity and vorticity fluctuations better than other models. RANS RST results showed a good agreement with the mean flow velocities, but was unable to capture flow fluctuations due its nature. RANS SST k-ω and DES SST k-ω models couldn't capture the jet flow separation region accurately. Earlier studies [34, 56] suggested that these methods may suffer from delayed estimation of reattachment.

A hybrid CAA method based on acoustic perturbation equation (APE) method proposed by Ewert and Schroder (2004) [41] was used to simulate the flow generated sound. CAA simulation was validated by comparing the simulated sound pressure spectra on the wall with the measured sound pressure spectra using a microphone. Sound sources and propagation were investigated by means of surface FFTs and POD analysis. The analysis showed that the peak frequency observed in sound pressure spectra is generated at the jet flow separation region downstream the stenosis. Analysis also showed that broadband frequencies were generated in the zone of high flow fluctuations and were damped significantly both downstream and upstream of this zone.

The validated numerical methods were then applied to study the flow generated sound in a realistic human lung airway model. High amplitude acoustic sources were observed near the anterior tracheal wall where the velocity jet generated at the glottis impinges on the tracheal wall. Simulations were run for different inspiratory flow rates to study the effect of flow rate on sound pressure level. Sound power increased with the flow rate while the rate of increase in sound power decreased with the increased in flow rate. These results were consistent with the results from previous experimental studies by Harper (2003)[55].

Future work will include development of experimental and simulation setup to study the sound generation and propagation in human lung airways to validate the applicability of proposed CFD and CAA methods to study breath sounds. Experimental setup will be developed to study the sound propagation and to localize sound sources using microphone arrays together with numerical techniques such as boundary element method (BEM) and beamforming [57] . The simulations will be enhanced by including the effects of flexible airway walls, airway secretions, and to simulate the sound propagation through surrounding tissues (including lung parenchyma and rib cage).

The validated numerical methods will also be performed in patient-specific airways with different pulmonary conditions. Numerical model of the patient-specific geometry could enable the prediction of breath sounds. Predicted sounds may then be compared with experimental measurements carried out on skin surface of the patients. The latter may be analyzed using signal processing algorithms [58-61] to filter noise (e.g. heart sound, environment noise, stomach sounds) before carrying out the comparison with computational results.  Furthermore, the extracted breath sound properties (such as time-frequency features) may be used to correlate these sounds with airway and lung pathology [62-74]. Later, classification of these sounds using machine learning (e.g. neural network, support vector machine) [75-78] would be carried out to detect pulmonary conditions.

# APPENDIX A: SURFACE FFT AND POD RESULTS

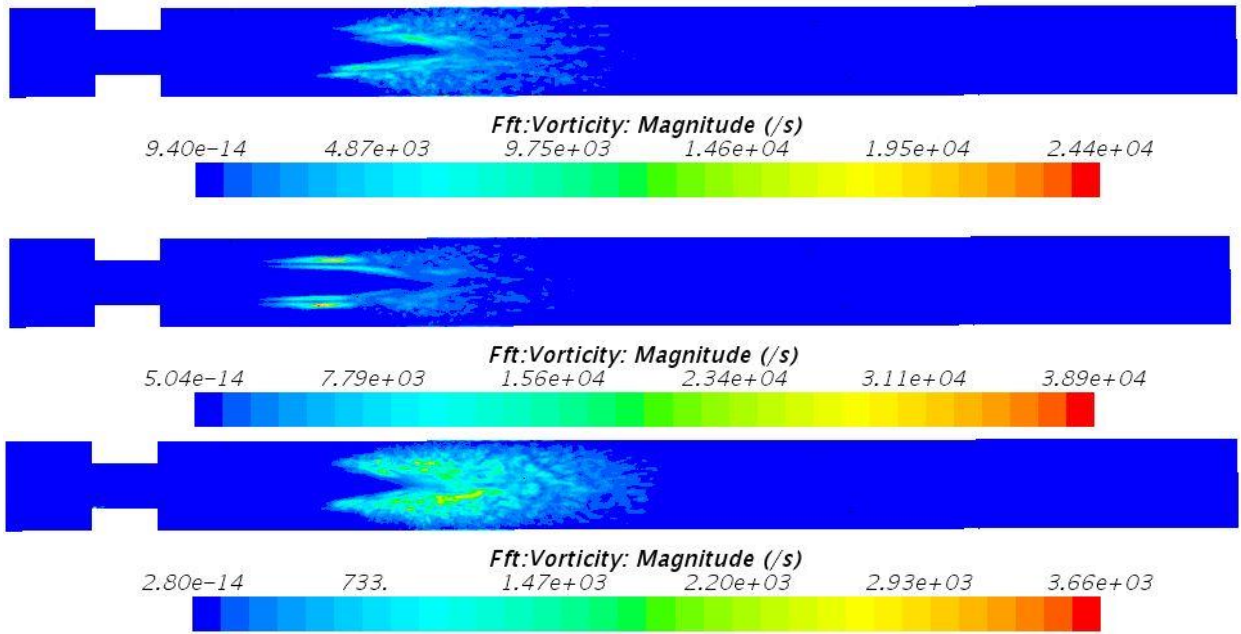

Figure 6-1: Surface FFTs of vorticity for frequencies 180Hz, 249 Hz, 420 Hz from top to bottom

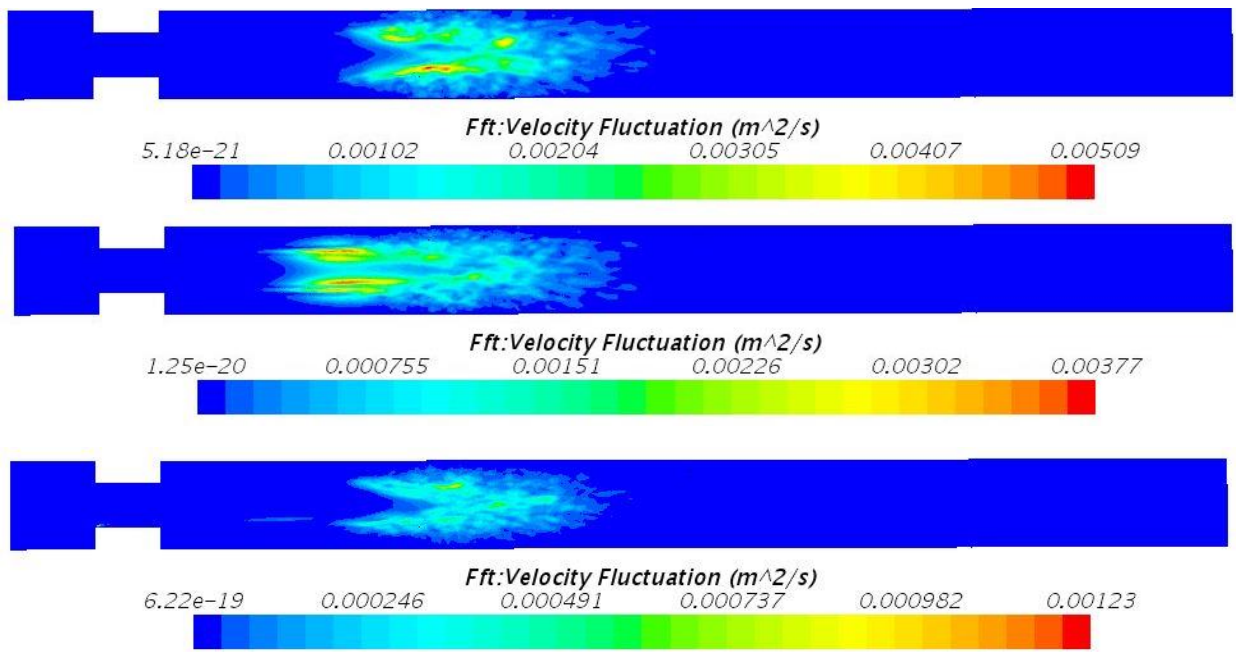

Figure 6-2 : Surface FFTs of velocity fluctuations for frequencies 180Hz, 249 Hz, 420 Hz from top to bottom

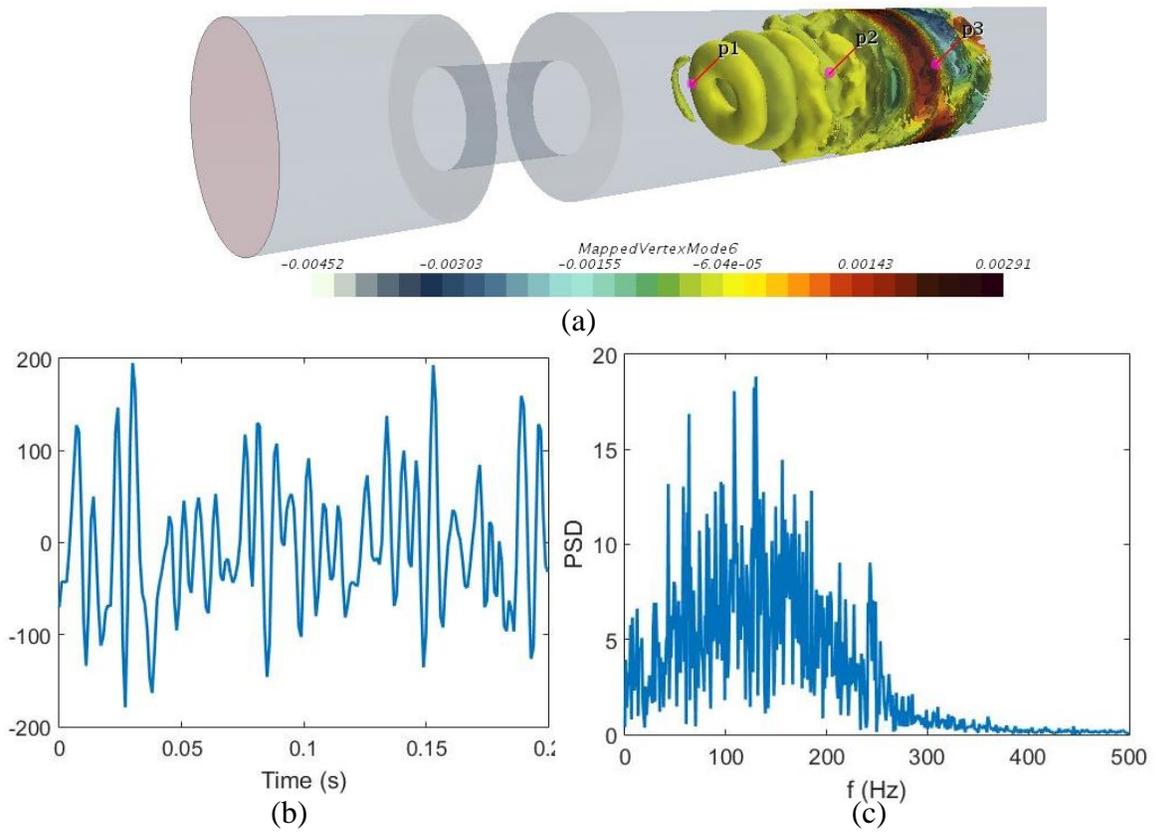

Figure 6-3 : Results for POD mode 6 of total sound pressure

# APPENDIX B: MATLAB CODE: POD

```matlab
clc
clear all
r=100; % number of modes
files = dir('*.csv');
n_snapshots = size(files,1); % number of snapshots

dt=0.0001;   % time step between two snapshts

for j=1:n_snapshots    % read data files from simulation
      fid = fopen(files(j).name,'r');

      data =  csvread(files(j).name,1,0);

      v(:,j)=data(:,1); %  Snapshot matrix

      fclose(fid);
end

U,S,V] = svd(v,0); % perform singular value decomposition of snapshot matrix

POD=U(:,1:r); % calculate r pod modes

PODT_coef=S*V'; % calculate POD time evolution matrix
```

[38]   J. F. Williams and D. L. Hawkings, "Sound generation by turbulence and surfaces in arbitrary motion," *Philosophical Transactions of the Royal Society of London A: Mathematical, Physical and Engineering Sciences,* vol. 264, pp. 321-342, 1969.
[39]   F. Khalili, P. Majumdar, and M. Zeyghami, "Far-Field Noise Prediction of Wind Turbines at Different Receivers and Wind Speeds: A Computational Study," in *ASME 2015 International Mechanical Engineering Congress and Exposition*, 2015, pp. V07BT09A051-V07BT09A051.
[40]   J. E. F. Williams and J. M. Lighthill, *Aerodynamic Generation of Sound*: Educational Development Center; distributor Encyclopaedia Britannica Educational Corporation, 1968.
[41]   R. Ewert and W. Schröder, "Acoustic perturbation equations based on flow decomposition via source filtering," *Journal of Computational Physics,* vol. 188, pp. 365-398, 2003.
[42]   J. F. Williams, "Hydrodynamic noise," *Annual Review of Fluid Mechanics,* vol. 1, pp. 197-222, 1969.
[43]   H. S. Ribner, "Aerodynamic sound from fluid dilitations; a theory of the sound from jets and other flows," University of Toronto1962.
[44]   M. Felli, S. Grizzi, and M. Falchi, "A novel approach for the isolation of the sound and pseudo-sound contributions from near-field pressure fluctuation measurements: analysis of the hydroacoustic and hydrodynamic perturbation in a propeller-rudder system," *Experiments in fluids,* vol. 55, p. 1651, 2014.
[45]   S. Grizzi and R. Camussi, "Wavelet analysis of near-field pressure fluctuations generated by a subsonic jet," *Journal of Fluid Mechanics,* vol. 698, pp. 93-124, 2012.
[46]   D. Violato and F. Scarano, "Three-dimensional vortex analysis and aeroacoustic source characterization of jet core breakdown," *Physics of fluids,* vol. 25, p. 015112, 2013.
[47]   J. Freund and T. Colonius, "Turbulence and sound-field POD analysis of a turbulent jet," *International Journal of Aeroacoustics,* vol. 8, pp. 337-354, 2009.
[48]   (2017). *Laser Doppler Anemometry Measurement Principles | Dantec Dynamics*. Available: https://www.dantecdynamics.com/measurement-principles-of-lda
[49]   G. Balafas, "Polyhedral mesh generation for CFD-analysis of complex structures," *Diplomityö. Münchenin teknillinen yliopisto,* 2014.
[50]   M. Ariff, S. M. Salim, and S. C. Cheah, "Wall y+ approach for dealing with turbulent flow over a surface mounted cube: part 1—low Reynolds number," in *Seventh International Conference on CFD in the Minerals and Process Industries*, 2009, pp. 1-6.
[51]   S. A. Ahmed and D. P. Giddens, "Flow disturbance measurements through a constricted tube at moderate Reynolds numbers," *Journal of biomechanics,* vol. 16, pp. 955-963, 1983.
[52]   M. K. Azad, H. A. Mansy, and P. T. Gamage, "Geometric features of pig airways using computed tomography," *Physiological reports,* vol. 4, p. e12995, 2016.
[53]   H. Y. Luo and Y. Liu, "Modeling the bifurcating flow in a CT-scanned human lung airway," *J Biomech,* vol. 41, pp. 2681-8, Aug 28 2008.
[54]   P. P. Thibotuwawa Gamage, F. Khalili, M. D. K. Azad, and H. A. Mansy, "Modeling Inspiratory Flow in a Porcine Lung Airway," *Journal of Biomechanical Engineering,* 2017.
81